\documentclass[12pt]{article}

\usepackage{amsfonts, amsmath, colonequals,amssymb}
\usepackage{algorithm}
\usepackage{multirow,colonequals}
\usepackage{mathtools,graphicx}
\usepackage{caption}
\usepackage{hyperref}
\usepackage{enumerate}
\usepackage{subcaption}
\usepackage{amsthm}
\usepackage{natbib}
\newcommand{\GG}[1]{}

\newcommand{\vecX}{\mathbf{X}}
\newcommand{\vecZ}{\mathbf{Z}}
\newcommand{\vecU}{\mathbf{U}}
\newcommand{\vecI}{\mathbf{I}}
\newcommand{\vecO}{\mathbf{O}}
\newcommand{\vecE}{\mathbf{E}}
\newcommand{\vecM}{\mathbf{M}}
\newcommand{\vecS}{\mathbf{S}}
\newcommand{\vecT}{\mathbf{T}}
\newcommand{\vecx}{\mathbf{x}}
\newcommand{\vecu}{\mathbf{u}}
\newcommand{\vecz}{\mathbf{z}}
\newcommand{\veczero}{\mathbf{0}}
\newcommand{\vecmu}{\mbox{\boldmath$\mu$}}
\newcommand{\veczeta}{\mbox{\boldmath$\zeta$}}
\newcommand{\vecalpha}{\mbox{\boldmath$\alpha$}}
\newcommand{\vecbeta}{\mbox{\boldmath$\beta$}}
\newcommand{\vecgamma}{\mbox{\boldmath$\gamma$}}

\newcommand{\vecLambda}{\mbox{\boldmath$\Lambda$}}

\newcommand{\matsig}{\mbox{\boldmath$\Sigma$}}
\newcommand{\matTheta}{\mbox{\boldmath$\Theta$}}
\newcommand{\matepsilon}{\mbox{\boldmath$\epsilon$}}
\newcommand{\matPsi}{\mbox{\boldmath$\Psi$}}
\newcommand{\varthet}{\mbox{\boldmath$\vartheta$}}
\newcommand{\ttr}{\text{tr}}

\newcommand{\isum}{\sum_{i=1}^n}
\newcommand{\gsum}{\sum_{g=1}^G}
\newcommand{\inv}{^{\raisebox{.2ex}{$\scriptscriptstyle-1$}}}
\DeclareMathOperator{\E}{\mathbb{E}}
\DeclareMathOperator{\R}{\mathbb{R}}

  \title{Flexible High-Dimensional Unsupervised\\ Learning with Missing Data}
  
\author{Yuhong Wei, Yang Tang and Paul D.\ McNicholas}
\date{\small Deptartment of Mathematics \& Statistics, McMaster University, Hamilton, Ontario, Canada.}

\begin{document}
\maketitle
\begin{abstract}
\noindent The mixture of factor analyzers (MFA) model is a famous mixture model-based approach for unsupervised learning with high-dimensional data. It can be useful, \textit{inter alia}, in situations where the data dimensionality far exceeds the number of observations. In recent years, the MFA model has been extended to non-Gaussian mixtures to account for clusters with heavier tail weight and/or asymmetry. The generalized hyperbolic factor analyzers (MGHFA) model is one such extension, which leads to a flexible modelling paradigm that accounts for both heavier tail weight and cluster asymmetry. In many practical applications, the occurrence of missing values often complicates data analyses. A generalization of the MGHFA is presented to accommodate missing values. Under a missing-at-random mechanism, we develop a computationally efficient alternating expectation conditional maximization algorithm for parameter estimation of the MGHFA model with different patterns of missing values. The imputation of missing values under an incomplete-data structure of MGHFA is also investigated. The performance of our proposed methodology is illustrated through the analysis of simulated and real data.\\[-6pt]

\end{abstract}

\section{Introduction}\label{sec:intro}

Model-based clustering is a popular exploratory analysis tool for unsupervised learning, or clustering. A finite mixture model is fitted to data, thereby revealing the group structure.  A finite mixture model is a convex linear combination of a finite number of component densities. Historically, the Gaussian mixture model has dominated the model-based clustering literature 
(e.g., \cite{celeux95,fraley02a}). However, the Gaussian mixture model is sensitive to both non-normality and the presence of heavy-tailed in the clusters. In recent years, finite mixtures of non-Gaussian distributions have flourished~(e.g., \cite{ browne15,lin16}). A recent review of model-based clustering is given by \cite{mcnicholas16b}, a review focusing on high-dimensional data is presented by \cite{bouveyron14}, and extensive details are given by \cite{mcnicholas16a}.

When clustering high-dimensional data where the number of variables $p$ is high relative to the number of observations $n$, model-based clustering techniques may produce unreliable results due to singular or near-singular estimates of the component covariance or scale matrices. In fact, larger values of $p$ alone can cause significant problems due to the fact that many mixture model-based approaches have $\mathcal{O}(p^2)$ free parameters. To introduce parsimony, families of mixture models have been developed by imposing constraints on the component covariance or scale matrices (e.g., \cite{celeux95, andrews12,vrbik14}). Each of these families arises via the imposition of constraints on the constituent parts of an eigen-decomposition of the component covariance or scale matrix (see \cite{banfield93}). Although these families of mixture models significantly reduce the number of free parameters in the component covariance or scale matrices, these matrices either remain $\mathcal{O}(p^2)$ or are diagonal. Accordingly, we either still have $\mathcal{O}(p^2)$ parameters in the component covariance or scale matrices or we have a model with very restrictive assumptions.

The mixture of factor analyzers (MFA) model (see \cite{ghahramani97}, \cite{mclachlan00}) reduces the number of model parameters to $\mathcal{O}(p)$. As the first robust modelling extension of MFA to accommodate atypical observations, \cite{andrews11} and \cite{mclachlan07} proposed mixtures of t-factor analyzers (MtFA). Since then, non-Gaussian analogues of mixtures of factor analyzers have gained popularity, including work on mixtures of skew-t factor analyzers (MSTFA; \cite{murray14a}), mixtures of skew-normal factor analyzers \cite{lin16}, mixtures of variance-gamma factor analyzers \cite{smcnicholas17}, and mixtures of generalized hyperbolic factor analyzers (MGHFA; \cite{tortora16}). The latter approach is particularly relevant to the work described herein.

Recently, more attention has been paid to the analysis of heterogeneous high-dimensional data involving different patterns of missing values. There are two strategies to convert a partially observed dataset to a completely observed one: deletion or imputation. Deletion removes the subjects with missing values, therefore  it is inadvisable when a substantial fraction of variables are affected. \cite{wagstaff05} propose a method that augments classical $k$-means clustering on deleted data via a tuning parameter for each variable containing missing entries based on the known relative importance of the variable in clustering. However, there are no guidelines on how to select the tuning parameters when the relative importance is unknown. Imputation fills in missing entries with plausible estimates of the missing values. Mean imputation and multiple imputation are two popular state-of-the-art frameworks for handling missing data (e.g. \cite{honaker11,su11,buuren10}). These imputation approaches work well only when the plausible values for the missing data can be identified. \cite{chi16}  propose the k-POD algorithm, which is a method for $k$-means clustering on partially observed data. The k-POD method employs a majorization-minimization (MM) algorithm (see \cite{hunter00},\cite{hunter04}) to identify a cluster that is in accord with the observed data. Because k-POD performs imputations iteratively, similar to the model-based clustering framework described herein, there are some similarities in how missing data are handled. However, the usual limitations of $k$-means apply to k-POD, e.g., it essentially fits spheres of equal radius.

Many model-based clustering techniques, such as the commonly used MFA and MtFA approaches, require complete data for statistical analysis. To overcome this weakness, \cite{wang13} generalized the mixture of common factor analyzers (MCFA) model --- which is more restrictive than the MFA model --- to accommodate missing values. To model high-dimensional data with heavier tailed clusters, \cite{wang15} further generalizes the mixture of common-t factor analyzers (MCtFA) approach to accommodating missing values. \cite{wei17} develop a mixture of generalized hyperbolic distributions and a mixture of skew-t distributions that account for missing data; however, these approaches are not applicable to high-dimensional data.

In this paper, we aim to develop a unified approach, based on the MGHFA model, for handling high-dimensional data in the presence of missing values as well as heavy-tailed and/or asymmetric clusters. Maximum likelihood estimates for our MGHFAMISS model are computed via a variant of the expectation-maximization (EM) algorithm \cite{dempster77}. Throughout, we assume that the data are missing-at-random (MAR; \cite{little87}), so that the missing data mechanism is ignorable. MAR means that the cause of the missingness is unrelated to the missing values, but may be related to the observed values of other variables. To ease the computational burden, two auxiliary permutation matrices are introduced, as in \cite{lin06}. As a by-product, the proposed procedure provides a conditional predictor to impute the missing values and a classifier to cluster partially observed vectors.

The remainder of the paper is organized as follows. In Section~\ref{sec:back}, we give a brief review of the generalized hyperbolic distribution and its building block, the generalized inverse Gaussian distribution. In Section~\ref{sec:meth}, we formulate the MGHFA model under an incomplete framework and study some of its statistical properties. Section~\ref{sec:mghdfamiss} describes the algorithm for parameter estimation and imputation of missing values via a conditional predictor. Some practical issues including the initial values and model selection are also addressed. In Section~\ref{sec:example}, the methodology is illustrated through simulated data with varying proportions of artificially missing values and a real ozone dataset with truly missing values. Finally, some concluding remarks are given in Section~\ref{sec:discussion}.

%%%%%%%%%%%%%%%
%%Review
%%%%%%%%%%%%%%%
\section{Background}\label{sec:back}
\subsection{Generalized Inverse Gaussian Distribution}

The random variable $W \in \R^{+}$ is said to have a generalized inverse Gaussian (GIG) distribution \cite{good53} with parameters $\lambda$, $\chi$, and $\psi$, denoted $W \sim \text{GIG}(\lambda,\chi,\psi)$, if its probability density function (pdf) is given by
\begin{equation}
\label{eqn:gig}
f_{\text{\tiny GIG}}(w;\lambda,\chi,\psi) = \frac{(\psi/\chi)^{\lambda/2}w^{\lambda-1}}{2K_\lambda(\sqrt{\psi\chi})}\text{exp}\left\{-\frac{\psi w+\chi/w}{2}\right\},
\end{equation}
where $\psi,\chi \in \R^{+}$, $\lambda \in \R$, and $K_\lambda(\cdot)$ is the modified Bessel function of the third kind with index $\lambda$.~\cite{barndorff77b},~\cite{blaesild78},~\cite{halgreen79}, and~\cite{jorgensen82} have demonstrated statistical properties of the GIG distribution, including the tractability of the following expectations:
\begin{align}
\label{eqn:expgig}
\nonumber
&\E [W] = \sqrt{\frac{\chi}{\psi}}\frac{K_{\lambda+1}(\sqrt{\psi\chi})}{K_{\lambda}(\sqrt{\psi\chi})},\qquad
\E [{1}/{W}] %= \sqrt{\frac{\psi}{\chi}}\frac{K_{\lambda-1}(\sqrt{\psi\chi})}{K_{\lambda}(\sqrt{\psi\chi})} 
= \sqrt{\frac{\psi}{\chi}}\frac{K_{\lambda+1}(\sqrt{\psi\chi})}{K_{\lambda}(\sqrt{\psi\chi})}-\frac{2\lambda}{\chi}, \\
\nonumber
&\E [\log W] = \log\left(\sqrt{\frac{\chi}{\psi}}\right)+\frac{\partial}{\partial\lambda}\log(K_\lambda(\sqrt{\psi\chi})).
\end{align}
These expected values lead to the development of a computationally efficient E-step for the parameter estimation that is presented in Section~\ref{sec:mghdfamiss}.

\cite{browne15} introduce an alternative parameterization of the GIG distribution by setting $\omega=\sqrt{\psi\chi}$ and $\eta=\sqrt{\chi/\psi}$. Write $W \sim \mathcal{I}(\lambda,\eta,\omega)$ to denote a random variable $W$ with this formulation and note that the density of $W$ is given by
\begin{equation}
f_{\mathcal{I}} (w\mid \lambda,\eta,\omega) = \frac{(w/\eta)^{\lambda-1}}{2\eta K_{\lambda}(\omega)} \text{exp} \left\{ -\frac{\omega}{2}\left(\frac{w}{\eta}+\frac{\eta}{w}\right) \right\},
\end{equation}
where $\eta \in \R^{+}$ is a scale parameter and $\omega \in \R^{+}$ is a concentration parameter. Note that this parameterization of the GIG distribution is an important ingredient for building the generalized hyperbolic distribution presented later.

\subsection{Multivariate Generalized Hyperbolic Distribution }
Several generalized hyperbolic distributions are available in the literature (e.g., \cite{browne15}, \cite{blaesild81}, \cite{mcneil05}). 
Following~\cite{browne15}, a $p\times 1$ random vector $\vecX$ is said to follow a generalized hyperbolic distribution, denoted by $\vecX \sim \text{GHD}_p(\lambda,\omega,\vecmu,\matsig,\vecbeta)$, if it can be represented by
\begin{equation}
\label{eqn:ghdstochastic}
\vecX = \vecmu + W\vecbeta + \sqrt{W}\vecU,
\end{equation}
$\vecU \bot W$, with index parameter $\lambda$, concentration parameter $\omega$, location vector $\vecmu$, dispersion matrix $\matsig$, and skewness vector $\vecbeta$. Here, $W \sim \mathcal{I}(\lambda,\eta=1,\omega)$, $\vecU \sim \mathcal{N}(\mathbf{0},\matsig)$, the symbol $\bot$ indicates independence, and it follows that $\vecX \mid w \sim \mathcal{N} (\vecmu + w\vecbeta, w\matsig)$. So, the pdf of the generalized hyperbolic random vector $\vecX$ is given by
\begin{equation*}\begin{split}
\label{eqn:ghdbrowne}
	f_{\text{\tiny GHD}}(\vecx &\mid \varthet) = \left [\frac{\omega +\delta(\vecx,\vecmu \mid \matsig )}{\omega+\vecbeta'\matsig\inv\vecbeta}\right]^{\frac{\lambda-p/2}{2}}\frac{K_{\lambda-p/2}\left(\sqrt{(\omega +\delta(\vecx,\vecmu \mid \matsig))(\omega+\vecbeta'\matsig\inv\vecbeta)}\right)}{(2\pi)^{p/2} |\matsig|^{1/2}K_\lambda(\omega)\text{exp}\{-(\vecx-\vecmu)'\matsig\inv\vecbeta\}},
\end{split}\end{equation*}
where $\delta (\vecx, \vecmu \mid \matsig)= (\vecx - \vecmu)'\matsig\inv(\vecx - \vecmu)$ is the squared Mahalanobis distance between $\vecx$ and $\vecmu$, $K_\lambda$ denotes the modified Bessel function of the third kind with index $\lambda$, and $\varthet= (\lambda, \omega, \vecmu, \matsig,\vecbeta)$ denotes the model parameters. 

%%%%%%%%%%%%%%%%%%%%%%
%MGHFA model formulation
%%%%%%%%%%%%%%%%%%%%%%
\section{Methodology}\label{sec:meth}
\subsection{MFA and MGHFA Models}\label{sec:mghfa}
%Out of consideration for completeness, we briefly outline the MFA and MGHFA models herein. The main idea behind MFA is to reduce the number of parameters in the specification of the component-covariance matrices. %Essentially, the mixture of factor analysis is combining factor analysis~\cite[FA;][]{spearman04} in the form of finite mixture~\cite{mclachlan00}. 
Given $n$ independent $p$-dimensional continuous variables $\vecX_1,\ldots,\vecX_n$, which come independently from a heterogeneous population with $G$ subgroups, the MFA model can be written as
\begin{equation}
\vecX_i = \vecmu_g + \vecLambda_g\vecU_{ig}+ \matepsilon_{ig}
\end{equation} 
with probability~$\pi_g$, for $i=1,\ldots,n$ and $g=1,\ldots,G$, where $\vecmu_g$ is a $p \times 1$ vector of component central location, $\vecLambda_g$ is a $p \times q$ matrix of  factor loadings, $\vecU_{ig} \sim \mathcal{N}(\veczero,\vecI_q)$ is a $q \times 1$ vector of latent factors, and $\matepsilon_{ig}\sim\mathcal{N}(\veczero,\matPsi_g)$ is a $p \times 1$ vector of errors with $\matPsi_g = \text{diag}(\psi_{g1},\ldots,\psi_{gp})$. Note that the $\vecU_{ig}$ are independently distributed and are independent of the $\matepsilon_{ig}$, which are also independently distributed. Under this  model, the marginal distribution of $\vecX_i$ from the $g$th component is $\mathcal{N}(\vecmu_g,\vecLambda_g\vecLambda_g'+\matPsi_g)$.

\cite{tortora16} consider an MGHFA model, where
\begin{equation}
\label{eqn:mghfastochastic}
\vecX_i = \vecmu_g + W_{ig}\vecbeta_g+\sqrt{W_{ig}}(\vecLambda_g\vecU_{ig}+ \matepsilon_{ig})
\end{equation}
with probability~$\pi_g$, where $W_{ig}\sim \mathcal{I}(\lambda_g,\eta=1,\omega_g)$, $\vecU_{ig} \sim \mathcal{N}(\veczero,\vecI_q)$, and $\matepsilon_{ig}\sim\mathcal{N}(\veczero,\matPsi_g)$. Note that $\vecU_{ig}$ and $\matepsilon_{ig}$ satisfy the same independence relationships as for the MFA model. It follows that
$\vecX_i\mid w_{ig}\sim\mathcal{N}(\vecmu_g+w_{ig}\vecbeta_g,w_{ig}(\vecLambda_g\vecLambda_g'+\matPsi_g))$. Then, they arrive at the MGHFA model with density
\begin{equation*}
g(\vecx \mid \varthet) = \gsum \pi_g f_{\tiny\text{GHD}}(\vecx \mid \lambda_g,\omega_g,\vecmu_g,\matsig_g,\vecbeta_g ),
\end{equation*}
where $\matsig_g = \vecLambda_g\vecLambda_g'+\matPsi_g$ and $\varthet$ denotes the model parameters. %$\vecpi=(\pi_1,\ldots,\pi_G)$. %The MGHFA model is a generalization of the MFA model by postulating a mixture of $G$ generalized hyperbolic factor analyzer sub-model for the distribution of $\vecX_i$. 

To denote which component each $\vecX_i$ belongs to, it is convenient to introduce %a set of membership labels $Z = (Z_1,\ldots,Z_n)$ with each 
$\vecZ_1,\ldots,\vecZ_n$, where $\vecZ_{i} = (Z_{i1},\ldots,Z_{iG})$ with $Z_{ig}=1$ if $\vecx_i$ belongs to the $g$th component and $Z_{ig} = 0$ otherwise. It follows that $\vecZ_i$ follows a multinomial distribution with one trial and cell probabilities $\pi_1,\ldots,\pi_G$, denoted by $\vecZ_i \sim \mathcal{M}(1; \pi_1,\ldots,\pi_G)$. From~\eqref{eqn:mghfastochastic}, a four-level hierarchical representation of MGHFA models can be formulated as follows:
\begin{align*}
%\label{eqn:four1}
\vecX_i \mid w_{ig}, \vecu_{ig}, z_{ig}=1 &\sim \mathcal{N}(\vecmu_g+w_{ig}\vecbeta_g+\vecLambda_g\vecu_{ig}, w_{ig}\matPsi_g),\\
%\label{eqn:four2}
\vecU_{ig} \mid w_{ig}, z_{ig} = 1 &\sim \mathcal{N}(\veczero,w_{ig}\vecI_q),\\
%\label{eqn:four3}
W_{ig} \mid z_{ig} = 1 &\sim \mathcal{I}(\lambda_g,\eta=1,\omega_g),\\
%\label{eqn:four4}
\vecZ_{i} &\sim \mathcal{M}(1;\pi_1,\ldots,\pi_G).
\end{align*}

\subsection{MGHFAMISS Model}
To set up updates for the MGHFAMISS model, $\vecX_i$ is partitioned into the observed component $\vecX_i^{\text{o}}$ and the missing component $\vecX_i^{\text{m}}$ with dimensions $p_{i}^{\text{o}} \times 1$ and $p_{i}^{\text{m}} \times 1$, respectively, where $p_{i}^{\text{o}}+p_{i}^{\text{m}}=p$. To facilitate computation, following~\cite{lin06}, indicator matrices are introduced, denoted by $\vecO_i~(p_{i}^{\text{o}} \times p)$ and $\vecM_i~(p_{i}^{\text{m}} \times p)$, which can be extracted from a $p$-dimensional identity matrix $\vecI_p$ corresponding to the respective row positions of $\vecX_i^{\text{o}}$ and $\vecX_i^{\text{m}}$ in $\vecX_i$, such that $\vecX_{i}^{\text{o}} = \vecO_i\vecX_i$ and $\vecX_{i}^{\text{m}} = \vecM_i\vecX_i$. It is not difficult to verify that $\vecX_i = \vecO_i'\vecX_i^{\text{o}} + \vecM_i'\vecX_i^{\text{m}}$ and $\vecO_i'\vecO_i + \vecM_i'\vecM_i = \vecI_p$. 
Now, some important consequences are summarized in the following proposition, which is useful for evaluating the required conditional expectation in the E-step of the algorithm described in the next section.

\textbf{Proposition 1} From the MGHFA model~\eqref{eqn:mghfastochastic} and the hierarchical representations given in Section~\ref{sec:mghfa}, we have:
\begin{list}{*}{\leftmargin=1em}
\item[a.] The conditional distribution of $\vecX_i^{\text{o}}$ given $w_{ig}$ and $z_{ig}=1$ is
		\begin{equation*}
		\vecX_i^{\text{o}} \mid w_{ig}, z_{ig}=1 \sim \mathcal{N}_{p_i^{\text{o}}}(\vecO_i(\vecmu_g+w_{ig}\vecbeta_g), w_{ig}\matsig_{ig}^{\text{oo}}),
		\end{equation*}
		where $\matsig_g = \vecLambda_g\vecLambda_g'+\matPsi_g$ and $\matsig_{ig}^{\text{oo}} = \vecO_i\matsig_g\vecO_i'$.
\item[b.] The marginal distribution of the observed component $\vecX_i^{\text{o}}$ is
		\begin{equation*}
		g(\vecx_i^{\text{o}}) = \gsum \pi_g f_{p_i^{\text{o}},\text{\tiny GHD}}(\vecx \mid \lambda_g,\omega_g,\vecmu_{ig}^{\text{o}},\matsig_{ig}^{\text{oo}},\vecalpha_{ig}^{\text{o}} ),
		\end{equation*}
		where $\vecmu_{ig}^{\text{o}} = \vecO_i\vecmu_g$, $\matsig_{ig}^{\text{oo}} = \vecO_i\matsig_g\vecO_i'$, $\vecalpha_{ig}^{\text{o}} = \vecO_i\vecbeta_g$, and $p_i^{\text{o}}$ is the dimension corresponding to the observed component  $\vecx_i^{\text{o}}$.
\item[c.] The conditional distribution of $\vecX_i^{\text{m}}$ given $\vecx_i^{\text{o}}$, $w_{ig}$, and $z_{ig}=1$ is
		\begin{equation*}
		\vecX_i^{\text{m}} \mid \vecx_i^{\text{o}}, w_{ig}, z_{ig}=1 \sim \mathcal{N}_{p_i^{\text{o}}}(\veczeta_{ig}^{\text{m}\cdot \text{o}},w_{ig}\matsig_{ig}^{\text{m}\cdot \text{o}}),
		\end{equation*}
where 
\begin{align*}
		\veczeta_{ig}^{\text{m}\cdot \text{o}} &= \vecM_i\left(\vecmu_g+w_{ig}\vecbeta_g+\matsig_g\vecS_{ig}^{\text{oo}}(\vecx_i-\vecmu_g-w_{ig}\vecbeta_g)\right),\\
		\matsig_{ig}^{\text{m}\cdot \text{o}} &= \vecM_i(\vecI_p-\matsig_g\vecS_{ig}^{\text{oo}})\matsig_g\vecM_i',\
		\vecS_{ig}^{\text{oo}} =\vecO_i'(\vecO_i\matsig_g\vecO_i')\inv\vecO_i.
		\end{align*}
\item[d.] We have %The conditional distribution of $W_{ig}$ given $\vecx_{i}^{\text{o}}$ and  $z_{ig}=1$ is
		\begin{equation}
		W_{ig} \mid \vecx_{i}^{\text{o}}, z_{ig}=1 \sim \text{GIG}(\lambda_{ig}^{\star},\chi_{ig}^{\star},\psi_{ig}^{\star}),
		\end{equation}
		where $\psi_{ig}^{\star} = \omega_g+\vecbeta_g\vecS_{ig}^{\text{oo}}\vecbeta_g'$, $\chi_{ig}^{\star} = \omega_g + (\vecx_i-\vecmu_g)'\vecS_{ig}^{\text{oo}}(\vecx_i-\vecmu_g)$, and $\lambda_{ig}^{\star}=\lambda_g - {p_{i}^{\text{o}}}/{2}$.
\item[e.] We have %The conditional distribution of $\vecX_i^{\text{o}}$ given $w_{ig}$, $\vecu_{ig}$, and $z_{ig}=1$ is
		\begin{equation}
		\vecX_i^{\text{o}} \mid w_{ig}, \vecu_{ig}, z_{ig}=1 \sim \mathcal{N}_{p_i^{\text{o}}}(\veczeta_{ig}^{\text{o}}, w_{ig}\matPsi_{ig}^{\text{oo}}),
		\end{equation}
		where $ \veczeta_{ig}^{\text{o}} = \vecO_i(\vecmu_g+w_{ig}\vecbeta_g+\vecLambda_g\vecu_{ig})$ and $\matPsi_{ig}^{\text{oo}} = \vecO_i\matPsi_g\vecO_i'$.
\item [f.] We have %The conditional distribution of $\vecX_i^{\text{m}}$ given $\vecx_i^{\text{o}}$, $w_{ig}$, $\vecu_{ig}$, and $z_{ig}=1$ is
		\begin{equation}
		\vecX_i^{\text{m}} \mid \vecx_i^{\text{o}}, w_{ig}, \vecu_{ig}, z_{ig}=1 \sim \mathcal{N}(\vecgamma_{ig}^{\text{m}\cdot \text{o}}, w_{ig} \matPsi_{ig}^{\text{m}\cdot \text{o}}),
		\end{equation}
where 
		\begin{align*}
		\vecgamma_{ig}^{\text{m}\cdot \text{o}} &= \vecM_i[\vecmu_g+w_{ig}\vecbeta_g+\vecLambda_g\vecu_{ig} +\matPsi_g\vecT_{ig}^{\text{oo}}(\vecx_i-\vecmu_g-w_{ig}\vecbeta_g-\vecLambda_g\vecu_{ig})],\\
		\matPsi_{ig}^{\text{m}\cdot \text{o}} &= \vecM_i(\vecI_p-\matPsi_g\vecT_{ig}^{\text{oo}})\matPsi_g\vecM_i',\qquad
		\vecT_{ig}^{\text{oo}} = \vecO_i'(\vecO_i\matPsi_g\vecO_i')\inv\vecO_i.
		\end{align*}
\item [g.] We have %The conditional distribution of $\vecU_{ig}$ given $\vecx_i^{\text{o}}$, $w_{ig}$, and $z_{ig}=1$ is
		\begin{equation*}\begin{split}
		\vecU_{ig} \mid & \vecx_i^{\text{o}}, w_{ig}, z_{ig}=1 \sim \mathcal{N}(\vecalpha_{ig}(\vecx_i-\vecmu_g-w_{ig}\vecbeta_g), w_{ig}(\vecI_q-\vecalpha_{ig}\vecLambda_g)),
		\end{split}\end{equation*}
		where $\vecalpha_{ig} = \vecLambda_g'\vecS_{ig}^{\text{oo}}$.
\end{list}
The proof of Proposition 1 is straightforward and hence omitted.

%%%%%%%%%%%%%%%%%%%%%%
%MGHFA with incomplete data
%%%%%%%%%%%%%%%%%%%%%%
\section{Computational Techniques}\label{sec:mghdfamiss}
\subsection{Learning %MGHFA Models With Incomplete Data 
via the AECM Algorithm}
To compute the maximum likelihood estimates for the parameters of MGHFA model with partially observed data, we adopt a modification of the expectation-conditional maximization (ECM) algorithm~\cite{meng93}, namely the alternating ECM (AECM) algorithm~\cite{meng97}. More precisely, the ECM  algorithm is an extension of the EM algorithm, where the M-step is simplified by performing a sequence of analytically tractable conditional maximization (CM) steps, and the AECM algorithm is an extension of the ECM algorithm where the specification of complete-data, i.e., the observed data plus the unobserved (missing and/or latent) data, is allowed to be different at each cycle of the algorithm. In our MGHFAMISS model, the complete-data is composed of the observed data $\vecx_i^{\text{o}}$ as well as the missing data $\vecx_i^{\text{m}}$, the missing labels $z_{ig}$, the latent $w_{ig}$, and the latent factors~$\vecu_{ig}$. 

For this application of the AECM algorithm to our MGHFAMISS model, one iteration consists of two cycles, with one E-step and five CM-steps in the first cycle and one E-step and two CM-steps in the second cycle. In the first cycle of the algorithm, we update the mixing proportions $\pi_g$, the component means $\vecmu_g$, the skewness $\vecbeta_g$, the concentration parameters $\omega_g$, and the index parameters $\lambda_g$. In the second cycle of the algorithm, we update the factor loadings matrices $\vecLambda_g$ and the error covariance matrices $\matPsi_g$. 

In the first cycle of the AECM algorithm, when estimating $\pi_g$, $\lambda_g$, $\omega_g$, $\vecmu_g$, and $\vecbeta_g$, the complete-data consist of the observed $\vecx_i^{\text{o}}$, the missing $\vecx_i^{\text{m}}$, the labels $z_{ig}$, and the latent $w_{ig}$. Hence, the complete-data log-likelihood is
\begin{equation}\label{eqn:logl1}\begin{split}
\log L_1 = \isum\gsum z_{ig}\big[&\log\pi_g +  \log\phi \left(\vecx_{i} \mid \vecmu_g + w_{ig} \vecbeta_g, w_{ig}\matsig_g\right)+\log h(w_{ig} \mid \omega_g, \lambda_g)\big].
\end{split}\end{equation}
%where $\phi \left(\vecx_i \mid \vecmu_g + w_{ig} \vecbeta_g, w_{ig}\matsig_g\right)$ is the Gaussian density with mean $\vecmu_g + w_{ig} \vecbeta_g$ and covariance $w_{ig}\matsig_g$.
In the E-step of the first cycle, in order to compute the expected value of the complete-data log-likelihood $\log L_1$, we need to compute $\E[Z_{ig}\mid\vecx_i^{\text{o}}]$, $\E[W_{ig}\mid\vecx_i^{\text{o}},z_{ig}=1]$, $\E[\log W_{ig}\mid\vecx_i^{\text{o}},z_{ig}=1]$, $\E[{1}/{W_{ig}}\mid \vecx_i^{\text{o}},z_{ig}=1]$, $\E[\vecX_i\mid\vecx_i^{\text{o}},z_{ig}=1]$, $\E[({1}/{W_{ig}})\vecX_i\mid\vecx_i^{\text{o}},z_{ig}=1]$, and $\E[({1}/{W_{ig}})\vecX_i\vecX_i'\mid\vecx_i^{\text{o}},z_{ig}=1]$.

As usual, the expected value of $Z_{ig}$ is given by
\begin{equation*}
\E[Z_{ig}\mid\vecx_i^{\text{o}}] = \frac{\pi_g f_{\text{GHD}}(\vecx_i^{\text{o}} \mid \lambda_g,\omega_g,\vecmu_{ig}^{\text{o}},\matsig_{ig}^{\text{oo}},\vecbeta_{ig}^{\text{o}})}{\sum_{h}^{G}\pi_h f_{\text{GHD}}(\vecx_i^{\text{o}} \mid \lambda_h,\omega_h,\vecmu_{ih}^{\text{o}},\matsig_{ih}^{\text{o}},\vecbeta_{ih}^{\text{o}})}\equalscolon\hat{z}_{ig}.
\end{equation*}
%where $\equalscolon$ means defined as. 
Let $a_{ig} = \E[W_{ig} \mid \vecx_{i}^{\text{o}}, z_{ig}=1]$, $b_{ig} = \E[1/W_{ig} \mid \vecx_{i}^{\text{o}}, z_{ig}=1]$, and $c_{ig} = \E[\log W_{ig} \mid \vecx_{i}^{\text{o}}, z_{ig}=1]$, which are implicit functions of parameters and can be evaluated directly by applying Propositions 1(d) and \eqref{eqn:expgig}. 

Recall that $\vecX_i = \vecO_i'\vecX_i^{\text{o}} + \vecM_i'\vecX_i^{\text{m}}$ and $\vecO_i'\vecO_i + \vecM_i'\vecM_i = \vecI_p$. These simply lead to $\vecO_i'\vecO_i(\vecI_p-\matsig_g\vecS_{ig}^{\text{oo}})=\veczero$. Then, based on Proposition 1(c), the following conditional expectations are obtained:
\begin{equation*}\begin{split}
\E&[\vecX_i\mid\vecx_i^{\text{o}},z_{ig}=1] =\vecmu_g+a_{ig}\vecbeta_g+\matsig_g\vecS_{ig}^{\text{oo}}(\vecx_i-\vecmu_g-a_{ig}\vecbeta_g) \equalscolon \vecE_{1ig}, \\
\E&[({1}/{W_{ig}})\vecX_i\mid\vecx_i^{\text{o}},z_{ig}=1] =b_{ig}\vecmu_g+\vecbeta_g+\matsig_g\vecS_{ig}^{\text{oo}}(b_{ig}(\vecx_i-\vecmu_g)-\vecbeta_g) \equalscolon \vecE_{2ig}, \\ 
\E&[({1}/{W_{ig}})\vecX_i\vecX_i'\mid\vecx_i^{\text{o}},z_{ig}=1] =(\vecI_p-\matsig_g\vecS_{ig}^{\text{oo}})\big[\matsig_g+(b_{ig}\vecmu_g\vecx_i'+\vecbeta_g\vecx_{i}')\vecS_{ig}^{\text{oo}}\matsig_g\\
&+(b_{ig}\vecmu_g\vecmu_g'+\vecmu_g\vecbeta_g'+\vecbeta_g\vecmu_g'+a_{ig}\vecbeta_g\vecbeta_g')(\vecI_p-\vecS_{ig}^{\text{oo}}\matsig_g)] +b_{ig}\matsig_g\vecS_{ig}^{\text{oo}}\vecx_i\vecx_i'\vecS_{ig}^{\text{oo}}\matsig_g\\ 
&+\matsig_g\vecS_{ig}^{\text{oo}}(b_{ig}\vecx_i\vecmu_g'+\vecx_{i}\vecbeta_g')(\vecI_p-\vecS_{ig}^{\text{oo}}\matsig_g)\equalscolon \vecE_{3ig}.
\end{split}\end{equation*}
After the expected value $Q_1$ of the complete-data log-likelihood \eqref{eqn:logl1} is formed, maximizing $Q_1$ with respect to $\pi_g$, $\vecmu_g$, and $\vecbeta_g$ gives rise to the parameter updates
\begin{align*}
&\hat{\pi}_g = \frac{n_g}{n}, \quad\hat{\vecmu}_g = \frac{\isum \hat{z}_{ig}(\bar{a}_g\vecE_{2ig}-\vecE_{1ig})}{\isum \hat{z}_{ig}(b_{ig}\bar{a}_g-1)}, \quad\text{and}\quad \hat{\vecbeta}_g =   \frac{\isum \hat{z}_{ig}(\bar{b}_g\vecE_{1ig}-\vecE_{2ig})}{\isum \hat{z}_{ig}(b_{ig}\bar{a}_g-1)},
\end{align*}
respectively, where $n_g = \isum \hat{z}_{ig}$, $\bar{a}_g = 1/n_g\isum \hat{z}_{ig}a_{ig}$, and $\bar{b}_g = 1/n_g\isum \hat{z}_{ig}b_{ig}$.  The estimates of the parameters $\omega_g$ and $\lambda_g$ are given as solutions to maximize the following function:
\begin{equation*}
q_g(\lambda_g,\omega_g) = -\log K_{\lambda_g}(\omega_g) + (\lambda_g-1)\bar{c}_g - \frac{\omega_g}{2}(\bar{a}_g+\bar{b}_g),
\end{equation*}
where $\bar{c}_g = 1/n_g\isum \hat{z}_{ig}c_{ig}$, and the associated updates are
\begin{align*}
\hat{\lambda}_g &= \bar{c}_g \hat{\lambda}_g^{\text{prev}}\left[\frac{\partial}{\partial \hat{\lambda}_{g}^{\text{prev}}} \log K_{\hat{\lambda}_g^{\text{prev}}} \left(\hat{\omega}_g^{\text{prev}}\right)\right]^{\inv},\\
\hat{\omega}_g &= \hat{\omega}_g^{\text{prev}} - \left[\frac{\partial}{\partial \hat{\omega}_g^{\text{prev}}} q_g \left(\hat{\omega}_g^{\text{prev}},\hat{\lambda}_g\right)\right]\left[\frac{\partial^2}{\partial (\hat{\omega}_g^{\text{prev}})^2} q_g \left(\hat{\omega}_g^{\text{prev}},\hat{\lambda}_g\right)\right]^{\inv},
\end{align*}
where the superscript `prev' denotes the previous estimate. %Note that these are analogous to the updates given by~\cite{browne15}.

In the second cycle of the AECM algorithm, when estimating $\vecLambda_g$ and $\matPsi_g$, the complete-data include the observed data $\vecx_i^{\text{o}}$, the missing data $\vecx_i^{\text{m}}$, the group labels $z_{ig}$, the latent $w_{ig}$, and the latent factors $\vecu_{ig}$. The complete-data log-likelihood can be written
\begin{equation}\label{eqn:l2}\begin{split}
\log L_2 =& \isum\gsum z_{ig}\bigr[\log\pi_g + \log\phi \left(\vecx_{i} \mid \vecmu_g + w_{ig} \vecbeta_g + \vecLambda_g\vecu_{ig}, w_{ig}\matPsi_g\right)\\
&\qquad\qquad\qquad\qquad+\log\phi (\vecu_{ig}\mid\veczero, w_{ig}\vecI_q)+\log h(w_{ig}\mid\omega_g, \lambda_g)\bigr],\\
%&\quad=C+\frac{1}{2}\isum\gsum z_{ig} \log |\matPsi_g^{\inv}|-\frac{1}{2}\isum\gsum z_{ig} \biggr[\ttr\left(\frac{1}{w_{ig}}(\vecx_i\vecx_i'-\vecx_i\vecmu_g'-\vecmu_g\vecx_i'+\vecmu_g\vecmu_g')\matPsi_g^{\inv}\right)\\
%&\qquad-2\ttr \left(\vecbeta_g(\vecx_i-\vecmu_g)'\matPsi_g^{\inv}\right)+\ttr \left(w_{ig}\vecbeta_g\vecbeta_g'\matPsi_g^{\inv}\right)-2\ttr \left(\frac{1}{w_{ig}}\matPsi_g^{\inv}\vecLambda_g\vecu_{ig}\vecx_i'\right)\\
%&\qquad+2\ttr \left(\frac{1}{w_{ig}}\vecmu_g'\matPsi_g'\vecLambda_g\vecu_{ig}\right)+2\ttr \left(\vecbeta_g'\matPsi_g^{\inv}\vecLambda_g\vecu_{ig}\right)+\ttr \left(\frac{1}{w_{ig}}\vecLambda_g\vecu_{ig}\vecu_{ig}'\vecLambda_g'\matPsi_g^{\inv}\right)\biggr],
\end{split}\end{equation}
%where $C$ is constant with respect to the parameters $\vecLambda_g$ and $\matPsi_g$. 
In the E-step of the second cycle, in order to compute the expected value of the complete-data log-likelihood $\log L_2$, in addtion to the same conditional expectations from the E-step of the first cycle, we will also need to compute $\E[\vecU_{ig}\mid\vecx_i^{\text{o}},z_{ig}=1]$, $\E[({1}/{W_{ig}})\vecU_i\mid\vecx_i^{\text{o}},z_{ig}=1]$, $\E[({1}/{W_{ig}})\vecU_i\vecU_i'\mid\vecx_i^{\text{o}},z_{ig}=1]$, and $\E[({1}/{W_{ig}})\vecU_i\vecX_i'\mid\vecx_i^{\text{o}},z_{ig}=1]$.

Recall that $\vecX_i = \vecO_i'\vecX_i^{\text{o}} + \vecM_i'\vecX_i^{\text{m}}$ and $\vecO_i'\vecO_i + \vecM_i'\vecM_i = \vecI_p$. These simply give rise to $\vecO_i'\vecO_i(\vecI_p-\matsig_g\vecS_{ig}^{\text{oo}})=\veczero$ and $\vecO_i'\vecO_i(\vecI_p-\matPsi_g\vecT_{ig}^{\text{oo}})=\veczero$. Then, based on Propositions 1(f) and 1(g), we obtain the following conditional expectations:
\begin{equation*}\begin{split}
\E&[\vecU_i\mid\vecx_i^{\text{o}},z_{ig}=1] =\vecalpha_{ig}(\vecx_i-\vecmu_g-a_{ig}\vecbeta_g) \equalscolon \vecE_{4ig}, \\
\E&[({1}/{W_{ig}})\vecU_i\mid\vecx_i^{\text{o}},z_{ig}=1] =\vecalpha_{ig}(b_{ig}(\vecx_i-\vecmu_g)-\vecbeta_g) \equalscolon \vecE_{5ig}, \\ 
\E&[({1}/{W_{ig}})\vecU_i\vecU_i'\mid\vecx_i^{\text{o}},z_{ig}=1] =\vecI_q-\vecalpha_{ig}\vecLambda_g+b_{ig}\vecalpha_{ig}(\vecx_i-\vecmu_g)(\vecx_i-\vecmu_g)'\vecalpha_{ig}'+a_{ig}\vecalpha_{ig}\vecbeta_g\vecbeta_g'\vecalpha_{ig}'\\&
\qquad-\vecalpha_{ig}\left((\vecx_i-\vecmu_g)\vecbeta_g'+\vecbeta_g(\vecx_i-\vecmu_g)'\right)\vecalpha_{ig}'\equalscolon \vecE_{6ig},\\
\E&[({1}/{W_{ig}})\vecU_i\vecX_i'\mid\vecx_i^{\text{o}},z_{ig}=1] = \vecE_{5ig}\vecx_i'\vecT_{ig}^{\text{oo}}\matPsi_g+\vecE_{5ig}\vecmu_g'(\vecI_p-\vecT_{ig}^{\text{oo}}\matPsi_g)+\vecE_{4ig}(\vecI_p-\vecT_{ig}^{\text{oo}}\matPsi_g)\\
&\qquad+\vecE_{6ig}\vecLambda_g'(\vecI_p-\vecT_{ig}^{\text{oo}}\matPsi_g)\equalscolon \vecE_{7ig}.
\end{split}\end{equation*}

Therefore, it follows that the expected value of the complete-data log-likelihood \eqref{eqn:l2} evaluated with $z_{ig}=\hat{z}_{ig}$, $\vecmu_g=\hat{\vecmu}_g$, and $\vecbeta_g=\hat{\vecbeta}_g$ is of the form
\begin{align*}
&Q_2 = \frac{1}{2}\isum\gsum \hat{z}_{ig} \log |\matPsi_g^{\inv}|-\frac{1}{2}\isum\gsum \hat{z}_{ig} \Bigl[\ttr\Big\{(\vecE_{3ig}-\vecE_{2ig}\hat{\vecmu}_g'
-\hat{\vecmu}_g\vecE_{2ig}'+b_{ig}\hat{\vecmu}_g\hat{\vecmu}_g')\matPsi_g^{\inv}\Big\}\\
&\qquad-2\ttr \Big\{\hat{\vecbeta}_g(\vecE_{1ig}-\hat{\vecmu}_g)'\matPsi_g^{\inv}\Big\}+\ttr \Big\{a_{ig}\hat{\vecbeta}_g\hat{\vecbeta}_g'\matPsi_g^{-1}\Big\}-2\ttr \Big\{\matPsi_g^{\inv}\vecLambda_g\vecE_{7ig}\Big\}+2\ttr \Big\{\hat{\vecmu}_g'\matPsi_g'\vecLambda_g\vecE_{5ig}\Big\}\\&\qquad +2\ttr \Big\{\hat{\vecbeta}_g'\matPsi_g^{\inv}\vecLambda_g\vecE_{4ig}\Big\}+\ttr \Big\{\vecLambda_g\vecE_{6ig}\vecLambda_g'\matPsi_g^{\inv}\Big\}\Bigr],
\end{align*}
%where the constant $C$ is omitted. 
ignoring terms that are constant with respect to $\vecLambda_g$ and $\matPsi_g$.
Differentiating $Q_2$ with respect to $\vecLambda_g$ and $\matPsi_g$, respectively, and setting the resulting derivatives equal to zero gives rise to their updates:
\begin{align*}
&\hat{\vecLambda}_g =\biggr[\isum \hat{z}_{ig}\left(\vecE_{7ig}'-\hat{\vecmu}_g\vecE_{5ig}'-\hat{\vecbeta}_g\vecE_{4ig}'\right)\biggr]\biggr[\isum \hat{z}_{ig}\vecE_{6ig}\biggr]^{\inv},\\
&\hat{\matPsi}_g =\frac{1}{n_g}\isum \hat{z}_{ig}\Big[\vecE_{3ig}-\vecE_{2ig}\hat{\vecmu}_g'-\hat{\vecmu}_g\vecE_{2ig}'+b_{ig}\hat{\vecmu}_g\hat{\vecmu}_g'
 -2\hat{\vecbeta}_g(\vecE_{1ig}-\hat{\vecmu}_g)'+a_{ig}\hat{\vecbeta}_g\hat{\vecbeta}_g'-2\hat{\vecLambda}_g\vecE_{7ig}\\
&\qquad+2\hat{\vecLambda}_g\vecE_{5ig}\hat{\vecmu}_g'+2\hat{\vecLambda}_g\vecE_{4ig}\hat{\vecbeta}_g'+\hat{\vecLambda}_g\vecE_{6ig}\hat{\vecLambda}_g'\Big].
\end{align*}

The AECM algorithm iteratively updates the parameters %after a certain number of iterations is reached or 
until a suitable convergence rule is satisfied. Herein, the Aitken acceleration~\cite{aitken26} was employed to stop our AECM algorithm. The Aitken acceleration at iteration $k$ is
$a^{(k)} = [{l^{(k+1)}-l^{(k)}}]/[{l^{(k)}-l^{(k-1)}}]$,
where $l^{(k)}$ is the log-likelihood value evaluated at iteration $(k)$. Following~\cite{bohning94}, an asymptotic estimate of the log-likelihood at iteration $k+1$ is given by
\begin{equation*}
l_\infty^{(k+1)} = l^{(k)} + \frac{1}{1-a^{(k)}}(l^{(k+1)}-l^{(k)}).\end{equation*}
\cite{mcnicholas10b} recommend that the AECM algorithm is stopped when $l_{\infty}^{(k+1)}-l^{(k)} < \epsilon$, provided that this difference is positive; we note that a similar criterion was proposed by \cite{lindsay95}. In the examples herein (Section~\ref{sec:example}), we set $\epsilon=10^{-5}$.

\subsection{Imputation of Missing Data}
When convergence is achieved, we obtain the maximum likelihood estimates of the parameters denoted by $\hat{\matTheta} = \{\hat{\pi}_g, \hat{\lambda}_g, \hat{\omega}_g, \hat{\vecmu}_g, \hat{\vecbeta}_g, \hat{\vecLambda}_g, \hat{\matPsi}_g: g = 1,\ldots, G\}$. Therefore, the \textit{a~posteriori} probability of group membership for each observation at convergence can be estimated by
\begin{equation*}\begin{split}
\text{P}(Z_{ig}=1\mid\vecx_i^{\text{o}};\hat{\matTheta})&=\frac{\hat{\pi}_g f_{\text{\tiny GHD}}(\vecx_i^{\text{o}} \mid \hat{\lambda}_g,\hat{\omega}_g,\hat{\vecmu}_{ig}^{\text{o}},\hat{\matsig}_{ig}^{\text{oo}},\hat{\vecbeta}_{ig}^{\text{o}})}{\sum_{h}^{G}\hat{\pi}_h f_{\text{\tiny GHD}}(\vecx_i^{\text{o}} \mid \hat{\lambda}_h,\hat{\omega}_h,\hat{\vecmu}_{ih}^{\text{o}},\hat{\matsig}_{ih}^{\text{o}},\hat{\vecbeta}_{ih}^{\text{o}})}\equalscolon\hat{z}_{ig}^{\star}.
\end{split}\end{equation*}
The resulting $\hat{z}_{ig}^{\star}$ can be used to cluster observations into groups based on the maximum \textit{a~posteriori} (MAP) probabilities. Specifically, $\text{MAP}(\hat{z}_{ig}^{\star})=1$ if $g=\arg\max_h(\hat{z}_{ih}^{\star})$ and $\text{MAP}(\hat{z}_{ig}^{\star}) = 0$ otherwise.

When analyzing incomplete data, it is often important to fill in the missing data with plausible values. We implement the imputation of the missing values based on the conditional mean method. That is, by substituting the maximum likelihood estimates $\hat{\vecmu}_g$, $\hat{\vecbeta}_g$, $\hat{\vecLambda}_g$, and $\hat{\matPsi}_g$ ($g = 1,\ldots, G$). This leads to a predictor of $\vecx_i^{\text{m}}$ given by
\begin{equation*}
%\vecx_i^{\text{m}} = 
\vecM_i \gsum \hat{z}_{ig}^{\star} (\hat{\vecmu}_g + {a}_{ig}\hat{\vecbeta}_g + \hat{\matsig}_g\hat{\vecS}_{ig}^{\text{oo}}(\vecx_i-\hat{\vecmu}_g-{a}_{ig}\hat{\vecbeta}_g)).
\end{equation*}

\subsection{Notes on implementation}
Similar to any EM-type iterative algorithm, the AECM algorithm may suffer from computational problems such as slow convergence or even failure to converge. %Indeed, our applications come with no guarantee of convergence to the global optimum when modelling multi-mode distributions in high-dimensions datasets that feature an overly large proportion of missing data. 
Often, good initial parameter values may speed up the convergence or lead to the attainment of a global optimum. To try to overcome computational difficulties, we recommend a simple procedure to automatically obtain a set of suitable initial values for the AECM algorithm, as follows.
\begin{list}{*}{\leftmargin=1em}
\item Perform mean imputation to fill in the missing values for each attribute separately, i.e., the missing value $\vecx_{ip}^{\text{m}}$ for the $i$th observation on the $p$th attribute is imputed by the sample mean of the observed values of the corresponding variable.
\item Perform $k$-means clustering to initialize the zero-one membership label $\hat{z}_{ig}^{(0)}$. Accordingly, the initial values for the model parameters are then
\begin{align*}
&\hat{\pi}_g^{(0)}=\frac{\isum \hat{z}_{ig}^{(0)}}{n},\quad\hat{\vecmu}_g^{(0)}=\frac{\isum\hat{z}_{ig}^{(0)}\vecx_i}{\isum \hat{z}_{ig}^{(0)}},\\
&\hat{\matsig}_g^{(0)}=\frac{\isum\hat{z}_{ig}^{(0)}(\vecx_i-\hat{\vecmu}_g^{(0)})(\vecx_i-\hat{\vecmu}_g^{(0)})'}{\isum \hat{z}_{ig}^{(0)}}.
\end{align*}
\item Generate the initial values for $\vecLambda_g$ and $\matPsi_g$ via the eigen-decomposition of $\hat{\matsig}_g^{(0)}$ as follows. The initial values of the $j$th column of $\vecLambda_g$ are set as $\gamma_{j}^{(0)} = \sqrt{d_j}\rho_{j}$, where $d_j$ is the $j$th largest eigenvalue of $\hat{\matsig}_g^{(0)}$ and $\rho_{j}$ is the $j$th eigenvector corresponding to the $j$th largest eigenvalue of $\hat{\matsig}_g^{(0)}$ for $j \in \{1, \ldots, q\}$. The matrix $\matPsi_g$ is then initialized as $\hat{\matPsi_g^{(0)}} = \text{diag}(\hat{\matsig}_g^{(0)}-\hat{\vecLambda}_g^{(0)}\hat{\vecLambda}_g^{(0)'})$.
\item Set the skewness parameter $\hat{\vecbeta}_g^{(0)} \approx \veczero$ for the near asymmetric assumption and set the index parameter $\hat{\lambda}_g^{(0)}=1$ and the concentration parameter $\hat{\omega}_g^{(0)}=-0.5$. %for simplicity.
\end{list}

To select an appropriate MGHFAMISS model in terms of the number of mixture components $G$ and the number of latent factors $q$, we adopt a widely used model selection criterion: the Bayesian information criterion (BIC; \cite{schwarz78}). The BIC is defined as
\begin{equation*}
\text{BIC} = 2l(\hat{\matTheta}) - \rho \log n,
\end{equation*}
where $l(\hat{\matTheta})$ is the maximized log-likelihood value, $\rho$ is the number of free parameters, and $n$ is the number of observations. 

While practical evidence~(e.g., \cite{mcnicholas08}, \cite{baek10}) suggests that the BIC performs well in choosing the number of mixture components and the number of latent factors, it is worthwhile to note that the BIC can be unreliable for the MFA models depending on the situation at hand (see \cite{baek11}, \cite{bhattacharya14}). Instead, \cite{baek11} suggest an alternative criterion to identify the suitable number of latent factors based on the approximate weight of evidence (AWE; \cite{banfield93}). The AWE is given by
\begin{equation*}
\text{AWE} = \text{BIC} - 2\text{EN}(\vecz)-\rho(3+\log n),
\end{equation*}
where $\text{EN}(\vecz_1,\ldots,\vecz_n) = -\isum\gsum \hat{z}_{ig}^{\star}\log \hat{z}_{ig}^{\star}$ is the entropy of the classification matrix with the $(i,g)$th entry being $\hat{z}_{ig}^{\star}$. Clearly, the AWE penalizes complex models more severely than the BIC, and thus tends to select more parsimonious models in practice. Bigger values of the BIC or AWE indicate preferable models. Nevertheless, there is no optimal strategy with respect to which criterion is the best, and a combined use of BIC and AWE may be helpful in selecting reasonable candidate models.

%%%%%%%%%%%%%%%%%%%
%Application 
%%%%%%%%%%%%%%%%%%%
\section{Numerical Examples}\label{sec:example}

\subsection{Simulation Studies}
To examine the performance of the MGHFAMISS model developed herein, we compare our proposed procedure to the existing mean imputation approach and the 
MSTFA model with missing values (MSTFAMISS). Respective EM algorithms for learning the MGHFAMISS and MSTFAMISS models are implemented in {\sf R} \cite{R16}. A two-step procedure is considered. First, the missing values are imputed according to mean imputation, where the missing 
values are replaced by their unconditional means. Next, the model parameters are estimated based on the ``completed'' data using some existing clustering methods found in {\sf R}, 
namely:
\begin{list}{*}{\leftmargin=1em}
\item Parsimonious Gaussian mixture models (PGMM; \cite{mcnicholas08}): model-based clustering using Gaussian mixtures of factor analyzers. We use the function {\tt{pgmmEM}} via the {\sf R} package {\tt{pgmm}} \cite{mcnicholas15} to derive the results. For the purpose of comparison, the covariance structure is set to be UUU, i.e., we fit the MFA model.
\item MGHFA \cite{tortora16}: model-based clustering using mixtures of generalized hyperbolic factor analyzers. The function {\tt{MGHFA}} via the {\sf R} package {\tt{MixGHD}} \cite{tortora15c} is used to derive the results. 
\end{list}  

\begin{figure}[!ht]
	\centering
~\hspace{-0.2in}\includegraphics[width=0.7\textwidth]{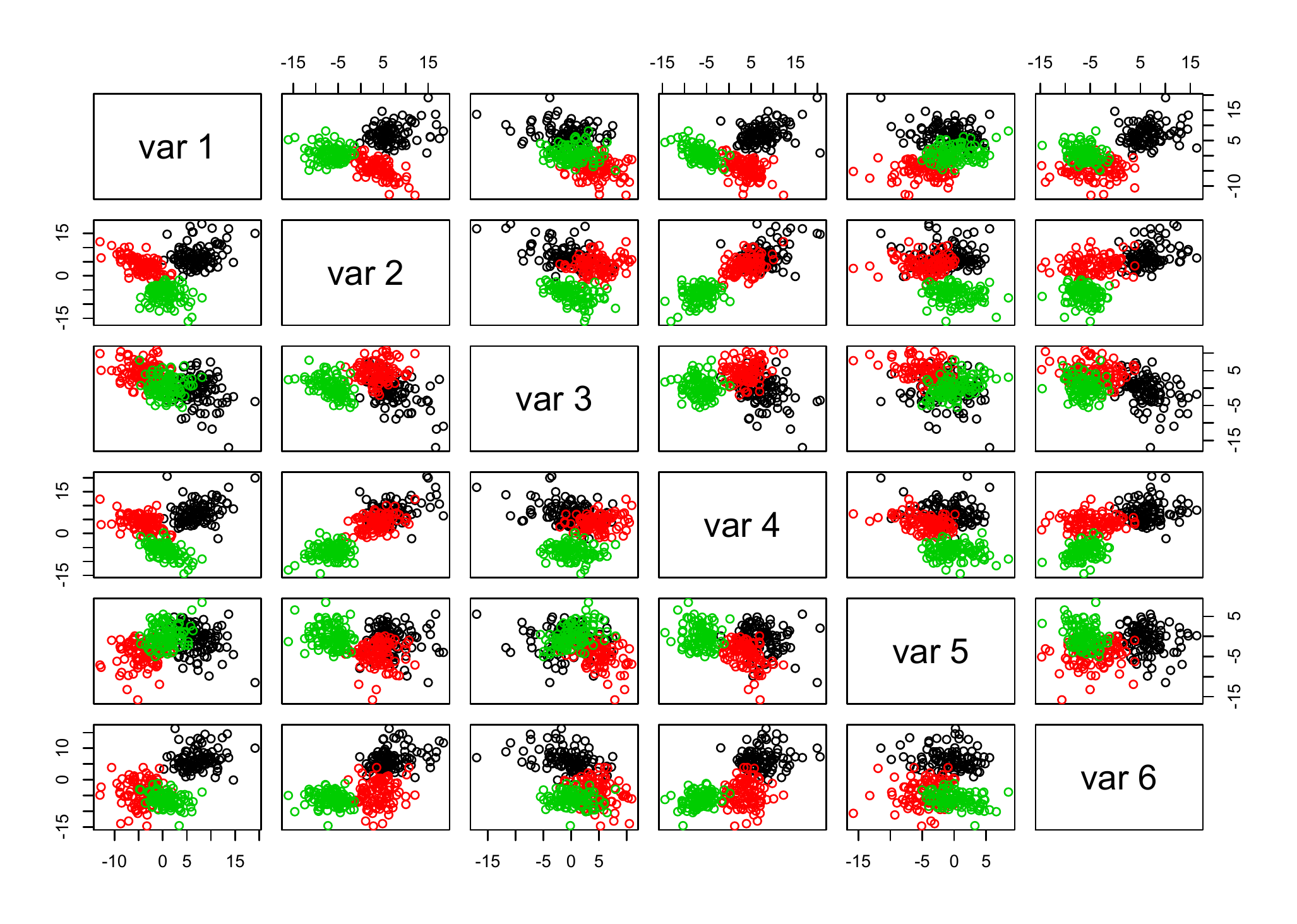}
	\vspace{-0.3in}
	\caption{Scatterplot of one of the simulated datasets, where colours reflect true class.}
	\label{fig:simudatapair}
\end{figure}

The samples were generated from a three-component MGHFA model with $q=2$ latent factors and $n_g=200$. Specifically, the data $\vecx_{i}$ were generated from 
\begin{equation}
\vecX_i = \vecmu_g + W_{ig}\vecbeta_g+\sqrt{W_{ig}}(\vecLambda_g\vecU_{ig}+ \matepsilon_{ig})
\end{equation}
with probability~$\pi_g$, where $\vecU_{ig}$ and $\matepsilon_{ig}$ satisfy distributional assumptions as in \eqref{eqn:mghfastochastic} and $g\in\{1,2,3\}$. The model parameters are given in Table~\ref{tab:simudata}. Figure~\ref{fig:simudatapair} depicts a scatterplot of the simulated data and its underlying clustering structure for one of the simulated datasets. 
%\vspace{-0.2in}
\begin{table*}[h]
	\caption{True model parameters for the simulated data.}
	\centering
	\begin{tabular*}{1.0\textwidth}{@{\extracolsep{\fill}}lll}
	\hline
	Component 1&Component 2 &Component 3\\
	\hline
	$\lambda_1=5$&$\lambda_2=3$&$\lambda_3=4$\\
	$\omega_1=3$&$\omega_2=6$&$\omega_3=6$\\
	$\vecmu_1 = (3,3,3,3,3,3)'$ &$\vecmu_2=(0,0,0,0,0,0)'$&$\vecmu_3 = (-3,-3,-3,-3,-3,-3)'$\\
	$\vecbeta_1=(1,1,-1,1,-1,1)$ &$\vecbeta_2 = (-1,1,1,1,1,-1,-1)'$&$\vecbeta_3 = (1,-1,1,-1,1,-1)'$\\
	$\vecLambda_1=\begin{pmatrix}-0.6&-0.1\\0.1&-0.5\\-0.8&0.8\\-0.6&-0.4\\0.1&-0.4\\0.8&-0.2\end{pmatrix}$&$\vecLambda_2=\begin{pmatrix}-0.5&-0.9\\0.4&1.0\\-0.5&-0.2\\-0.4&0.4\\0.5&0.3\\-0.8&0.9\end{pmatrix}$&$\vecLambda_3=\begin{pmatrix}0.7&-0.4\\0.8&0.0\\-0.2&0.9\\-0.3&0.4\\0.3&0.7\\-0.8&0.1\end{pmatrix}$\\
$\matPsi_1= 2\vecI_6 $&$\matPsi_2 = \vecI_6$&$\matPsi_3 = \vecI_6 $\\
	\hline
	\end{tabular*}
	\label{tab:simudata}
\end{table*}

Synthetic missing datasets are simulated by removing $n\times r$ elements from each column through three different MAR patterns under four missing rates: $r=5\%$, $r=10\%$, $r=20\%$ and $r=30\%$. Data points in each column $c$ ($c=1,\ldots 5$) are sorted in descending order. Column $c+1$ is then divided into three equal blocks and, for each block, a specified number of elements (see Table~\ref{tab:pat}) are removed at random. When $c=6$, the first column is used rather than column $c+1$.

\begin{table*}[h]
	\caption{Number of missing observations for each pattern.}
	\centering
	\begin{tabular*}{1.0\textwidth}{@{\extracolsep{\fill}}llll}
	\hline
	$r$&Pattern 1&Pattern 2 &Pattern 3\\
	\hline
$5\%$&(4,20,6)&(20,4,6)&(6,4,20)\\
$10\%$&(8,40,12)&(40,8,12)&(12,8,40)\\
$20\%$&(16,80,24)&(80,16,24)&(24,16,80)\\
$30\%$&(24,120,36)&(120,24,36)&(36,24,120)\\
\hline
	\end{tabular*}
	\label{tab:pat}
\end{table*}

For comparison, group memberships are initialized using $k$-means clustering. The clustering experiments comprise 30 replications per combination of missing pattern and missingness rate. The performance assessments in terms of classification are evaluated through the adjusted Rand index (ARI; \cite{hubert85}) and misclassification (error) rates (ERR). In this study, we fit the simulated data using PGMM with mean imputation (MI-PGMM), MGHFA with mean imputation (MI-MGHFA), MSTFAMISS, and MGHFAMISS models with $G=3$ and $q=2$.

Tables~\ref{tab:pat1},~\ref{tab:pat2} and~\ref{tab:pat3} report the mean of the BIC, AWE, ARI, and ERR together with their corresponding standard deviations (Std. Dev.) under each combination considered. Moreover, the frequencies (Freq.) supported by the BIC and AWE are also recorded. Not surprisingly, the results indicate that the best model based on the BIC and AWE is an MGHFAMISS model. At low levels of missingness (i.e., $r=5\%$ and $r=10\%$), all methods perform well for all three patterns. The performance drops significantly for MI-PGMM and MI-MGHFA at the highest level of missingness (i.e., $r=30\%$). Moreover, the ARI values from mean imputation approaches are different for each pattern when $r=30\%$. Both MGHFAMISS and MSTFAMISS perform well at high levels of missingness, giving much higher ARI and much lower ERR than those resulting from the MI-PGMM and MI-MGHFA models. 

\begin{table*}[!ht]
	\caption{Simulation results based on 30 replications for missing pattern 1.}
	\centering
	\footnotesize{
	\begin{tabular*}{1.0\textwidth}{@{\extracolsep{\fill}}llllll}
	  \hline
	  Criteria &  & MI-PGMM & MI-MGHFA & MSTFAMISS & MGHFAMISS \\ 
	  \hline
	  $r=5\%$ &  &  &  &  &  \\ 
	  \hline
	  & Mean & -18847 & -8030 &-7055 & -7026 \\ 
	  BIC & Std. Dev. & 43 & 128&69  & 62 \\ 
	  & Freq. &  &  &  & 30 \\ 
	  & Mean &  &  &-7958  & -7928 \\ 
	  AWE & Std. Dev. &  &&70  & 91\\ 
	  & Freq. &  & &  & 30 \\ 
	  \multirow{2}{*}{ARI}& Mean &0.97 & 0.95 & 0.99 & 0.99 \\ 
	  & Std. Dev. & 0.00&  0.12& 0.01 & 0.00\\ 
 	 \multirow{2}{*}{ERR}& Mean &0.01 &0.03  &0.00 & 0.01 \\ 
 	 & Std. Dev. &0.00 & 0.08&0.00 & 0.00 \\ 
 	 \hline
	  $r=10\%$ &  &  &  &  &  \\ 
	  \hline
	  & Mean &-19023 &-8281 &-6866&-6782\\ 
	  BIC & Std. Dev. &67&120&97&68 \\ 
	  & Freq. &  &  &  & 30 \\ 
	  & Mean &  &&-7877&-7695 \\ 
	  AWE & Std. Dev.& &&99&69  \\ 
	  & Freq. &  & & 1 & 29 \\ 
	  \multirow{2}{*}{ARI}& Mean &0.94&0.93&0.98&0.98 \\ 
	  & Std. Dev. &0.01 &0.13&0.01&0.01\\ 
	  \multirow{2}{*}{ERR}& Mean &0.02&0.03&0.01&0.01 \\ 
	  & Std. Dev. &0.00 &0.07 &0.00&0.00\\ 
	  \hline
	  $r=20\%$ &  &  &  &  &  \\ 
	  \hline
	  & Mean &-19163&-8662 &-6249&-6228\\ 
	  BIC & Std. Dev. &64 &131&64&60\\ 
 	 & Freq. &  &  & 4 & 26 \\ 
 	 & Mean &  &&-7191&-7169 \\ 
	  AWE & Std. Dev. &  & &67&63 \\ 
	  & Freq. & &  & 5 & 25 \\ 
	  \multirow{2}{*}{ARI}& Mean &0.83 &0.86&0.95&0.95 \\ 
	  & Std. Dev. & 0.01 &0.12&0.01&0.02\\ 
	  \multirow{2}{*}{ERR}& Mean &0.06 &0.05 &0.02&0.02\\ 
	  & Std. Dev. &0.01 &0.08&0.01&0.01\\ 
	  \hline
	  $r=30\%$ &  &  &  &  &  \\ 
	  \hline
	  & Mean &-19055&-8828&-5745&-5654\\ 
 	 BIC & Std. Dev. &68&171 &60&58 \\ 
 	 & Freq. &  &  & 6 & 24 \\ 
 	 & Mean &  & &-6673&-6647 \\ 
 	 AWE & Std. Dev. &&&65&66 \\ 
 	 & Freq. &  &  & 3 & 27 \\ 
 	 \multirow{2}{*}{ARI}& Mean & 0.32&0.69&0.89&0.90\\ 
 	 & Std. Dev. &0.18 &0.20&0.01&0.02\\ 
	  \multirow{2}{*}{ERR}& Mean &0.29 &0.14&0.04&0.04 \\ 
	  & Std. Dev. &0.10 &0.14 &0.00&0.01\\ 
	   \hline
	\end{tabular*}}
	\label{tab:pat1}
\end{table*}

\begin{table*}[!ht]
	\caption{Simulation results based on 30 replications for missing pattern 2.}
	\centering
	\footnotesize{
	\begin{tabular*}{1.0\textwidth}{@{\extracolsep{\fill}}llllll}
	  \hline
	  Criteria &  & MI-PGMM & MI-MGHFA & MSTFAMISS & MGHFAMISS \\ 
	  \hline
	  $r=5\%$ &  &  &  &  &  \\ 
	  \hline
	  & Mean &-18905&-8131&-7044&-7038\\ 
	  BIC & Std. Dev. &47&124&99&72  \\ 
	  & Freq. &  &  &  & 30 \\ 
	  & Mean &  &  &  -7947& -7940 \\ 
	  AWE & Std. Dev. &  &&101  &72 \\ 
	  & Freq. &  & &  & 30 \\ 
	  \multirow{2}{*}{ARI}& Mean &0.96 &0.91&0.99&0.99\\ 
	  & Std. Dev. &0.00&0.16&0.01&0.01\\ 
 	 \multirow{2}{*}{ERR}& Mean &0.01&0.05 &0.00&0.00\\ 
 	 & Std. Dev. &0.00&0.11&0.00&0.00 \\ 
 	 \hline
	  $r=10\%$ &  &  &  &  &  \\ 
	  \hline
	  & Mean &-19078&-8434&-6796&-6771  \\ 
	  BIC & Std. Dev. &64 &108&98&85\\ 
	  & Freq. &  &  & 1 & 29 \\ 
	  & Mean &  &&-7707&-7682 \\ 
	  AWE & Std. Dev. & &&100&86 \\ 
	  & Freq. &  & & 1 & 29 \\ 
	  \multirow{2}{*}{ARI}& Mean & 0.92&0.92&0.98&0.98\\ 
	  & Std. Dev. & 0.01&0.14&0.01&0.01\\ 
	  \multirow{2}{*}{ERR}& Mean & 0.03&0.04&0.01&0.01\\ 
	  & Std. Dev. &0.00 &0.09&0.00&0.00 \\ 
	  \hline
	  $r=20\%$ &  &  &  &  &  \\ 
	  \hline
	  & Mean &-19180&-8925&-6219&-6215 \\ 
	  BIC & Std. Dev. &66&89&85&94 \\ 
 	 & Freq. &  &  & 5 & 25 \\ 
 	 & Mean &  &&-7160&-7155 \\ 
	  AWE & Std. Dev. &  & &87&99 \\ 
	  & Freq. & &  & 5 & 25 \\ 
	  \multirow{2}{*}{ARI}& Mean & 0.77&0.88&0.96&0.95 \\ 
	  & Std. Dev. & 0.04 &0.02&0.01&0.02\\ 
	  \multirow{2}{*}{ERR}& Mean &0.08  &0.04&0.02&0.02\\ 
	  & Std. Dev. &0.03&0.01&0.00&0.01 \\ 
	  \hline
	  $r=30\%$ &  &  &  &  &  \\ 
	  \hline
	  & Mean &-18749&-9232&-5759&-5708\\ 
 	 BIC & Std. Dev. &62& 80.24 &88&64\\ 
 	 & Freq. &  &  & 6 & 24 \\ 
 	 & Mean &  & &-6790&-6709 \\ 
 	 AWE & Std. Dev. &&&74&67 \\ 
 	 & Freq. &  &  & 4 & 26 \\ 
 	 \multirow{2}{*}{ARI}& Mean & 0.15&0.66&0.88&0.89\\ 
 	 & Std. Dev. &0.13&0.16&0.01&0.03 \\ 
	  \multirow{2}{*}{ERR}& Mean &  0.45&0.14&0.04&0.04\\ 
	  & Std. Dev. & 0.19&0.12&0.00&0.01 \\ 
	   \hline
	\end{tabular*}}
	\label{tab:pat2}
\end{table*}

\begin{table*}[!ht]
	\caption{Simulation results based on 30 replications for missing pattern 3.}
	\centering
	\footnotesize{
	\begin{tabular*}{1.0\textwidth}{@{\extracolsep{\fill}}llllll}
	  \hline
	  Criteria &  & MI-PGMM & MI-MGHFA & MSTFAMISS & MGHFAMISS \\ 
	  \hline
	  $r=5\%$ &  &  &  &  &  \\ 
	  \hline
	  & Mean & -18898 &-8027&-7074&-7066\\ 
	  BIC & Std. Dev. &68&144&87&89\\ 
	  & Freq. &  &  &  & 30 \\ 
	  & Mean &  &  &  -7989&-7969  \\ 
	  AWE & Std. Dev. &  &&  92&90 \\ 
	  & Freq. &  & &  & 30 \\ 
	  \multirow{2}{*}{ARI}& Mean &0.96&0.90&0.99&0.99 \\ 
	  & Std. Dev. &0.01&0.19&0.01&0.01\\ 
 	 \multirow{2}{*}{ERR}& Mean & 0.01&0.05&0.00&0.00\\ 
 	 & Std. Dev. & 0.00&0.11&0.00&0.00\\ 
 	 \hline
	  $r=10\%$ &  &  &  &  &  \\ 
	  \hline
	  & Mean & -19097 &-8279&-6795&-6771\\ 
	  BIC & Std. Dev. &67&92 &85&87\\ 
	  & Freq. &  &  & 1 & 29 \\ 
	  & Mean &  &&-7708&-7682 \\ 
	  AWE & Std. Dev. & &&87&88 \\ 
	  & Freq. &  & & 1 & 29 \\ 
	  \multirow{2}{*}{ARI}& Mean & 0.93&0.95&0.98&0.98\\ 
	  & Std. Dev. &0.03&0.02&0.01&0.01 \\ 
	  \multirow{2}{*}{ERR}& Mean &0.03&0.02 &0.01&0.01\\ 
	  & Std. Dev. &0.01 &0.01 &0.00&0.00\\ 
	  \hline
	  $r=20\%$ &  &  &  &  &  \\ 
	  \hline
	  & Mean &-19229&-8713&-6250&-6244 \\ 
	  BIC & Std. Dev. & 72&115&80&83\\ 
 	 & Freq. & 0&  & 6 & 24 \\ 
 	 & Mean &  & &-7192&-7186\\ 
	  AWE & Std. Dev. &  & &83&89 \\ 
	  & Freq. & &  & 5 & 25 \\ 
	  \multirow{2}{*}{ARI}& Mean &0.56 &0.87&0.95&0.94 \\ 
	  & Std. Dev. & 0.23&0.11&0.02&0.02 \\ 
	  \multirow{2}{*}{ERR}& Mean & 0.17 &0.05&0.02&0.02\\ 
	  & Std. Dev. & 0.11&0.06&0.01&0.01\\ 
	  \hline
	  $r=30\%$ &  &  &  &  &  \\ 
	  \hline
	  & Mean &-19062&-9478&-5644&-5636\\ 
 	 BIC & Std. Dev. &69&289&84&37  \\ 
 	 & Freq. &  &  & 8 & 22 \\ 
 	 & Mean &  & &-6639&-6627 \\ 
 	 AWE & Std. Dev. &&&84&35 \\ 
 	 & Freq. &  &  & 3 & 27 \\ 
 	 \multirow{2}{*}{ARI}& Mean &0.18&0.59&0.88&0.89 \\ 
 	 & Std. Dev. & 0.27&0.29&0.02&0.02\\ 
	  \multirow{2}{*}{ERR}& Mean & 0.42&0.21 &0.04&0.04\\ 
	  & Std. Dev. & 0.20 &0.21&0.01&0.01\\ 
	   \hline
	\end{tabular*}}
	\label{tab:pat3}
\end{table*}

Next, the predictive accuracy of the imputation of missing values is explored. The empirical discrepancy measure for imputed values is simply
\begin{equation*}
\text{MSE} =  \frac{1}{n^*}\sum_{i=1}^{n}(\vecx_i^{\text{m}}-\hat{\vecx}_i^{\text{m}})'(\vecx_i^{\text{m}}-\hat{\vecx}_i^{\text{m}}),
\end{equation*}
where $n^*=\sum_{i=1}^n (c-c_i^{\text{o}})$ is the number of missing values. Table~\ref{tab:simumse} shows the mean MSE together with its standard deviations. The MGHFAMISS and MSTFAMISS models substantially outperform MI for all cases. 
\begin{table*}[!ht]
	\caption{Imputation performance for MI-PGMM, MI-MGHFA, MGHFAMISS, and MSTFAMISS models under various missing rates ($r$) for Pattern 1}
	\centering
	\begin{tabular*}{1.0\textwidth}{@{\extracolsep{\fill}}llrrrr}
	  \hline
	  &  & \multicolumn{4}{c}{MSE}  \\ 
	  \cline{3-6}
	  $r$ &  & MI-PGMM & MI-MGHFA & MSTFAMISS & MGHFAMISS \\ 
	 	 \hline
	  \multirow{2}{*}{5\%} & Mean & 30.14 & 30.14 & 8.87& 8.70 \\ 
	  & Std. Dev. & 3.01 &3.01 & 1.43 & 1.43 \\ 
	  \multirow{2}{*}{10\%} & Mean & 30.14 & 30.14 & 8.96 & 8.97 \\ 
	  & Std. Dev. & 3.08 & 3.08 & 0.93 & 0.89 \\ 
	  \multirow{2}{*}{20\%} & Mean & 29.15 & 29.15 & 9.58 & 9.78 \\ 
	  & Std. Dev. & 1.75 & 1.75 & 0.91 & 0.91 \\ 
	  \multirow{2}{*}{30\%} & Mean & 28.91 & 28.91 & 10.87 & 10.88 \\ 
	  & Std. Dev. & 1.47 & 1.47 & 0.78 & 0.80 \\ 
	   \hline
	\end{tabular*}
	\label{tab:simumse}
\end{table*}

We then compare our approach with the k-POD algorithm \cite{chi16}, via the function \texttt{kpod} in the {\sf R} package {\tt kpodclustr}. Table~\ref{tab:kpod} reports the mean of the ARI and ERR together with their corresponding standard deviations (Std. Dev.) under various missing rates for Pattern 1. The MGHFAMISS approach substantially outperforms k-POD in all cases with the presence of  longer tails and asymmetry in data. Notably, [22] show their result is superior to that of state-of-the-art imputation methods, such as \texttt{Amelia} imputation \cite{honaker11}, \texttt{mi} imputation \cite{su11} and \texttt{mice} imputation \cite{buuren10}. 
\begin{table*}[!ht]
	\caption{Simulation results based on 30 replications using MGHFAMISS and k-POD for Pattern 1}
	\centering
	{\small\begin{tabular*}{1.05\textwidth}{@{\extracolsep{\fill}}llrrrrrrrr}
	  \hline	
	&&\multicolumn{4}{c}{MGHFAMISS} &\multicolumn{4}{c}{k-POD}\\ \cline{3-6} \cline{7-10}
	  &&$r=5\%$    & $r=10\%$ & $r=20\%$ & $r=30\%$&$r=5\%$    & $r=10\%$ & $r=20\%$ & $r=30\%$ \\ 
	\hline
	  \multirow{2}{*} {ARI}&Mean&0.99&0.98&0.95&0.90&0.92&0.87 &0.75&0.62 \\
	&Std.Dev.&0.00&0.01&0.02&0.02&0.02&0.02&0.03&0.06\\
	    \multirow{2}{*} {ERR}&Mean&0.01&0.01&0.02&0.04 &0.03 &0.04&0.09&0.14\\
	&Std.Dev.&0.00&0.00&0.01&0.04&0.01&0.01&0.01&0.04\\	 
	   \hline
	\end{tabular*}}
	\label{tab:kpod}
\end{table*}

\begin{figure}[!ht]
	\centering
~\hspace{-0.05in}\includegraphics[width=0.7\textwidth]{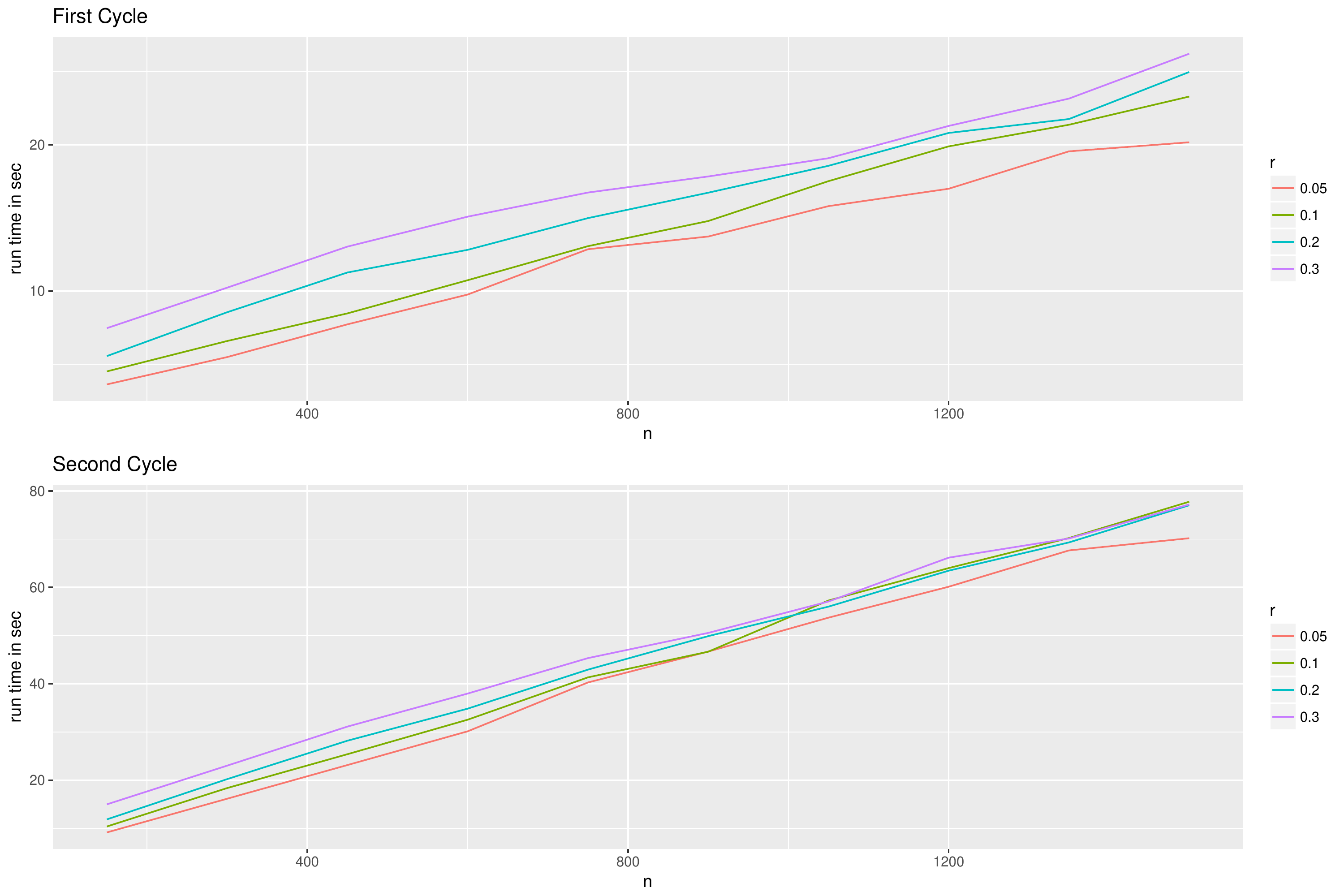}
	\vspace{-0.2in}
	\caption{Plot of run time (in seconds) over 100 repetions under various $n$ and $r$.}
	\label{fig:runtime}
\end{figure}

To explore the speed of the proposed algorithm, we generate samples with $n\in\{150, 300, \ldots, 1500\}$ under various missing rates for Pattern 1. Table~\ref{tab:runtime} and Figure~\ref{fig:runtime} show the run time (in seconds) per iteration over 100 repetitions of the experiment. We see that the run time increases linearly with the sample size $n$ for both cycles. Figure~\ref{fig:runtime} shows that the missing rate has an impact on run time for the first cycle only. 

 \begin{table*}[!ht]
	\caption{Run time (in seconds) over 100 repetions under various $n$ and $r$.}
	\centering
	{\scriptsize\begin{tabular*}{1.0\textwidth}{@{\extracolsep{\fill}}lrrrrrrrr}
	  \hline
	   & \multicolumn{2}{c}{$r=5\%$}& \multicolumn{2}{c}{$r=10\%$ }& \multicolumn{2}{c}{$r=20\%$}&\multicolumn{2}{c} {$r=30\%$ }\\ \cline{2-3}\cline{4-5}\cline{6-7}\cline{8-9}
	$n$ &$1^{\text{st}}$ Cycle&$2^{\text{nd}}$ Cycle&$1^{\text{st}}$ Cycle&$2^{\text{nd}}$ Cycle&$1^{\text{st}}$ Cycle&$2^{\text{nd}}$ Cycle&$1^{\text{st}}$ Cycle&$2^{\text{nd}}$ Cycle\\  
	\hline  
	150&3.61&9.18&4.50&10.41&5.55&11.88&7.46&15.00\\
	300&5.48&16.17&6.58&18.38&8.55&20.23&10.23&23.03\\
	450&7.72&23.15&8.47&25.40&11.27&28.20&13.04&31.13\\
	600&9.76&30.15&10.74&32.56&12.82&34.85&15.09&37.98\\
	750&12.86&40.28&13.07&41.34&14.99&42.94&16.74&45.33\\
	900&13.73&46.69&14.79&46.66&16.73&49.89&17.84&50.56\\
	1050&15.81&53.72&17.52&57.28&18.57&56.01&19.09&57.07\\
	1200&17.00&60.14&19.90&64.01&20.82&63.47&21.30&66.17\\
	1350&19.56&67.67&21.38&70.24&21.77&69.33&23.17&73.15\\
	1500&20.18&70.20&23.31&77.77&24.99&77.04&26.24&80.20\\
	\hline		
		\end{tabular*}}
	\label{tab:runtime}
\end{table*}

\subsection{Italian Wine Data}
In addition to the simulated data experiments, our MGHFAMISS approach is applied to real data. In this first experiment, we apply MGHFAMISS to the well-known Italian wine data, collected by~\cite{forina86} on wines grown in the same region in Italy but derived from three cultivars: 59 Barolo, 71 Grignolino, and 48 Barbera. There are $n=178$ samples of $p=13$ physical and chemical features available in the {\tt{gclus}} package~\cite{hurley04} for {\sf R}.

The wine data are standardized prior to analysis using the {\tt{scale}} function in {\sf R}. Then, we modify the normalized wine data by adding seventeen noisy attributes, which are irrelevant for clustering purposes, to the original attributes. Following \cite{wang13}, the noise attributes are generated from an independent uniform distribution on the interval $(-1,1)$. These two datasets (i.e., original wine data and modified wine data) are complete, so for illustration purposes we remove entries through an MAR mechanism to obtain approximately 5\%, 10\%, 20\%, and 30\% overall missingness.

To compare the BIC and the AWE with respect to choosing the number of latent factors, the MGHFAMISS model with $g=3$ and $q=1,\ldots,7$ are applied for parameter estimation. Simulations were run with a total of thirty replications under each scenario considered.

Table~\ref{tab:winefreq} summarizes the frequencies of each of the candidate models preferred by the BIC and the AWE for the original and modified wine data under various missing rates. Similar to \cite{wang13}, the AWE tends to select models with a smaller number of factors than BIC does.
Compared to \cite{wang13}, our proposed MGHFA model chooses a smaller number of latent factors based on BIC and the same number of latent factors based on AWE.

Table~\ref{tab:winearierr} lists averaged ARI and mean ERR together with their corresponding standard deviations under the MGHFAMISS and the MSTFAMISS models. As anticipated, as the missingness rates increase, the ARI values and the ERR values generally decrease and increase, respectively. Adding noisy variables leads to a slight worsening of the classification performance. In addition, the averaged ARI under the MGHFAMISS models is higher than the MSTFAMISS models except for the highest level of missingness  (i.e., $r=30\%$). This is not surprising because the clusters in the wine data are not highly skewed. However, when the missing rate reaches $30\%$, the two approaches yield similar results.
 
\begin{table*}[!ht]
\caption{The frequencies with which each of the MGHFAMISS models (run for $q=1,\ldots,7$) are chosen by the BIC and AWE for the original and modified wine data under various missingness rates; frequencies are $0$ for $q>3$ and so are omitted.}
\centering
{\scriptsize\begin{tabular*}{1.06\textwidth}{@{\extracolsep{\fill}}lrrrrrrrrrrrrrrrrr}
  \hline
  & \multicolumn{8}{c}{Original wine data} && \multicolumn{8}{c} {Modified wine data} \\ 
  \cline{2-9}\cline{11-18}
   & \multicolumn{2}{c}{5\%}&\multicolumn{2}{c}{10\%}&\multicolumn{2}{c}{20\%}&\multicolumn{2}{c}{30\%}&&\multicolumn{2}{c}{5\%}&\multicolumn{2}{c}{10\%} & \multicolumn{2}{c}{20\%} &\multicolumn{2}{c}{30\%} \\ 
   \cline{2-3}\cline{4-5}\cline{6-7}\cline{8-9}\cline{11-12}\cline{13-14}\cline{15-16}\cline{17-18}
$q$ & BIC & AWE & BIC & AWE & BIC & AWE & BIC & AWE 
  && BIC & AWE & BIC & AWE & BIC & AWE & BIC & AWE \\ 
  \hline
  1 & 16 & 30 & 24 & 30 & 29 & 30 & 30 & 30 
     && 30 & 30 & 30 & 30 & 30 & 30 & 30 & 30 \\ 
  2 & 14 & 0 & 4 & 0 & 1 & 0 & 0 & 0 
     && 0 & 0 & 0 & 0 & 0 & 0 & 0 & 0 \\ 
  3 & 0 & 0 & 2 & 0 & 0 & 0 & 0 & 0 
     && 0 & 0 & 0 & 0 & 0 & 0 & 0 & 0 \\ 
%  4--7 & 0 & 0 & 0 & 0 & 0 & 0 & 0 & 0 
 %        && 0 & 0 & 0 & 0 & 0 & 0 & 0 & 0 \\ 
%  5 & 0 & 0 & 0 & 0 & 0 & 0 & 0 & 0 & 0 & 0 & 0 & 0 & 0 & 0 & 0 & 0 \\ 
%  6 & 0 & 0 & 0 & 0 & 0 & 0 & 0 & 0 & 0 & 0 & 0 & 0 & 0 & 0 & 0 & 0 \\ 
%  7 & 0 & 0 & 0 & 0 & 0 & 0 & 0 & 0 & 0 & 0 & 0 & 0 & 0 & 0 & 0 & 0 \\ 
   \hline
\end{tabular*}}
\label{tab:winefreq}
\end{table*}

\begin{table*}[!ht]
\caption{The averaged ARI and ERR values for the best MGHFAMISS and MSTFAMISS models based on BIC for the original and modified wine data under various missingness rates.}
\centering
\begin{tabular*}{1.0\textwidth}{@{\extracolsep{\fill}}lrrrrrrrrr}
  \hline
&\multicolumn{4}{c}{Original wine data}&\multicolumn{4}{c}{Modified wine data}\\
\cline{2-5} \cline{6-9}
 & \multicolumn{2}{c}{MGHFAMISS} & \multicolumn{2}{c}{MSTFAMISS} &  \multicolumn{2}{c}{MGHFAMISS} &  \multicolumn{2}{c}{MSTFAMISS}\\
\cline{2-3} \cline{4-5} \cline{6-7} \cline{8-9}
$r$ & ARI & ERR & ARI & ERR & ARI & ERR & ARI & ERR \\ 
  \hline 
  \multirow{2}{*}{5} & 0.85 & 0.05 & 0.82 & 0.06 & 0.87 & 0.04 & 0.82 & 0.06 \\ 
  & (0.08) & (0.06) & (0.06) & (0.02) & (0.06) & (0.02) & (0.05) & (0.02) \\ 
  \multirow{2}{*}{10} & 0.82 & 0.06 & 0.78 & 0.08 & 0.82 & 0.06 & 0.75 & 0.06 \\ 
  & (0.08) & (0.06) & (0.07) & (0.03) & (0.05) & (0.02) & (0.08) & (0.03) \\ 
 \multirow{2}{*}{20} & 0.77 & 0.08 & 0.75 & 0.09 & 0.72 & 0.08 & 0.70 & 0.08 \\ 
  & (0.07) & (0.03) & (0.10) & (0.09) & (0.22) & (0.07) & (0.20) & (0.03) \\ 
 \multirow{2}{*}{30} & 0.75 & 0.09 & 0.76 & 0.08 & 0.72 & 0.07 & 0.72 & 0.08 \\ 
  & (0.08) & (0.03) & (0.08) & (0.06) & (0.21) & (0.03) & (0.21) & (0.03) \\ 
\hline
\end{tabular*}
\label{tab:winearierr}
\end{table*}

\subsection{Ozone Level Detection Data}
To further demonstrate the proposed methodology, ozone level detection data with truly missing values are analyzed herein. The dataset, available from the UCI Machine Learning Repository~\cite{lichman13}, was originally collected by \cite{zhang06} for the Houston, Galveston, and Briazoria (HGB) area from several databases within two major federal data warehouses and one local database for air quality control. These are, respectively, the United States Environmental Protection Agency Air Quality System and National Climate Data Center from the federal government and Continuous Ambient Monitoring Stations operated by the Texas Commission on Environmental Quality. There are two ground ozone level datasets: one is the one hour peak set, the other is the eight hour peak set, and both consist of at least 2500 observations with 72 continuous features containing various measures of air pollutant and meteorological information for the HGB area. As stated by~\cite{zhang08}, forecasting ozone days is challenging because the dataset
is sparse, contains a large number of irrelevant features (only about 10 out of 72 features have been verified by environmental scientists to be useful and relevant), has (cluster) skewness, and has a lot of missing values.

The one hour ozone data feature 73 ozone days versus 2463 normal days and the eight hour ozone data feature 160 ozone days versus 2374 normal days. Both datasets contain 8.2\% missing values. The status of whether a day is an ozone day or normal day was recorded for each observation, and is naturally used as the true class variable. These datasets have been previously analyzed by \cite{wang13} and \cite{zhang08}. \cite{wang13} analyzed these datasets using an MCFA approach.

Before performing the fitting, we scale the partially observed dataset using the {\tt{scale}} function in {\sf R}. Following~\cite{wang13}, we fit a two-component MGHFAMISS model with $q= 1,\ldots,60$. Note that the largest number of latent factors is chosen such that the relationship $(p-q)^2>(p+q)$
is satisfied (see \cite{lawley62}).
\begin{figure*}[!ht]
	\centering
	\includegraphics[width=0.8\textwidth]{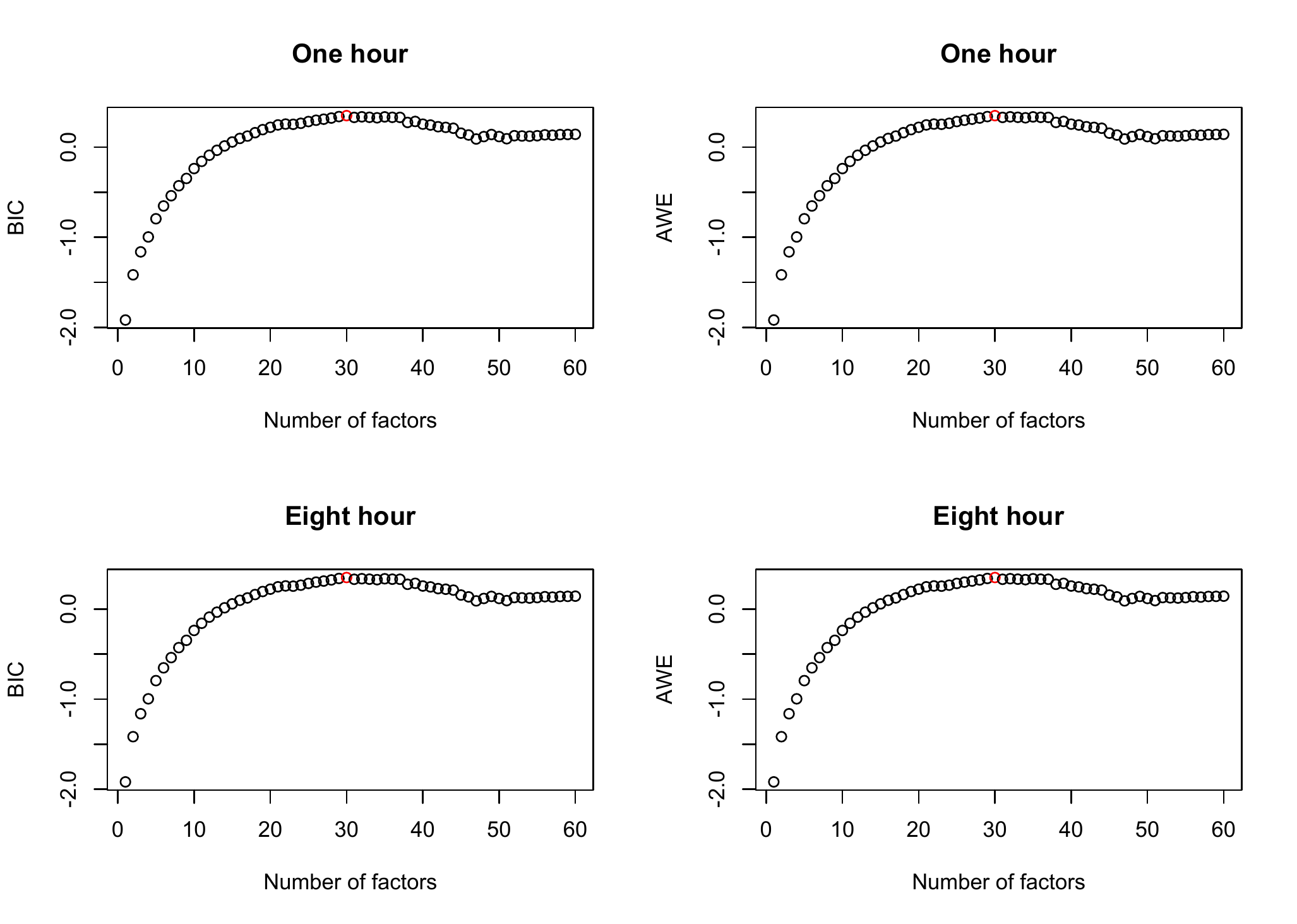}
	\caption{Plot of BIC and AWE values versus number of latent factors $q$ for the MGHFAMISS models fitted to the one hour and eight hour ozone data, where the maximum is highlighted in each case}
	\label{fig:ozonebicaweversusq}
\end{figure*}

Considering a plot of the BIC and AWE values versus the number of latent factors for the MGHFAMISS model (Figure~\ref{fig:ozonebicaweversusq}), the BIC and the AWE both prefer $q=30$ for the one and eight hour ozone data. The best model reported by~\cite{wang13} had $q=43$ and $q=44$, based on the BIC, for the one hour and eight hour ozone data, respectively, and $q=34$, based on the AWE, for both datasets. 
\cite{zhang08} points out that there are a large number of irrelevant features for both datasets; accordingly, it is notable that our MGHFAMISS approach prefers smaller $q$ when compared to \cite{wang13}.

To assess the classification performance, following ~\cite{wang13}, we apply 7-fold (in terms of years) cross-validation (CV) procedures and estimate the correct classification rate (i.e., $1-\text{ERR}$) for both the one hour and eight hour ozone data. Observations from one of the seven years are treated as the testing data and the remaining observations are treated as training data. The correct classification rate lies in the range from 57.9\% to 71.7\% and from 54.6\% to 73.2\% for the one hour and eight hour ozone data, respectively. Even though the classification accuracy is not very high, it is slightly superior to the maximum correct classification rate of 72.5\% reported by~\cite{wang13} for the eight hour ozone data. Notably, they show their result is superior to that of the GMIX imputation~\cite{lin06} and the {\tt{mclust}}~\cite{fraley12} methods. %\begin{color}{blue}which usually encounter a convergence problem in most cases\end{color}. 

%%%%%%%%%%%%%%%%%%%
%Discussion and Conclusion
%%%%%%%%%%%%%%%%%%%
\section{Discussion}\label{sec:discussion}

The MGHFA model has been extended to accommodate complex missing patterns for high-dimensional data with heavy tails and strong asymmetry. By borrowing the attractive features of the GIG distribution, we developed an efficient and elegant parameter estimation for the MGHFAMISS model within an AECM framework. To simplify matrix manipulations, two auxiliary permutation matrices were incorporated in the procedure. %The proposed AECM algorithm can simultaneously take into account the missing values and clustering purpose. 
The analysis of simulated and real data reveal that the proposed method is quite effective for the reconstruction of the missing values and outperforms other competing models for unsupervised learning when data contain missing information and clusters exhibit non-normal features such as asymmetry and/or heavy tails. The wine data example shows the MGHFAMISS model can be superior to the MSTFAMISS model when the data has a relatively low missingness rate and clusters that are not highly skewed. 

There are computational challenges that must be addressed when fitting the MGHFAMISS model. Most particularly, the AECM algorithm requires the imputation of missing values on each iteration of the algorithm and, as the number of missing values becomes large, this task becomes increasingly time consuming. Implementing this approach in parallel %computing via the {\tt{parallel}} package in the {\sf R} code can 
would help to ease this computational burden. Also, families of parsimonious models could be obtained by considering a generalized hyperbolic analogue to the PGMM models of \cite{mcnicholas08} and \cite{mcnicholas10}. Future work will also include investigation of alternatives to the AECM algorithm for parameter estimation, e.g., via a Bayesian approach (e.g., \cite{utsugi01}, \cite{lin04}, \cite{lin09b}). Alternatives to the BIC and the AWE for selecting the number of latent factors $q$, such as the LASSO-penalized BIC \cite{bhattacharya14}, will be considered for model selection.

\section*{Acknowledgments}

The authors gratefully acknowledge financial support from the Ontario Graduate Scholarship, the Faculty of Science Research Fellows, and Canada Research Chairs programs.


\begin{thebibliography}{}

\bibitem[\protect\citeauthoryear{Aitken}{Aitken}{1926}]{aitken26}
Aitken, A.~C. (1926).
\newblock A series formula for the roots of algebraic and transcendental
  equations.
\newblock {\em Proceedings of the Royal Society of Edinburgh\/}~{\em 45\/}(1),
  14--22.

\bibitem[\protect\citeauthoryear{Andrews and McNicholas}{Andrews and
  McNicholas}{2011}]{andrews11}
Andrews, J.~L. and P.~D. McNicholas (2011).
\newblock Extending mixtures of multivariate t-factor analyzers.
\newblock {\em Statistics and Computing\/}~{\em 21\/}(3), 361--373.

\bibitem[\protect\citeauthoryear{Andrews and McNicholas}{Andrews and
  McNicholas}{2012}]{andrews12}
Andrews, J.~L. and P.~D. McNicholas (2012).
\newblock Model-based clustering, classification, and discriminant analysis via
  mixtures of multivariate t-distributions.
\newblock {\em Statistics and Computing\/}~{\em 22\/}(5), 1021--1029.

\bibitem[\protect\citeauthoryear{Baek and McLachlan}{Baek and
  McLachlan}{2011}]{baek11}
Baek, J. and G.~J. McLachlan (2011).
\newblock Mixtures of common t-factor analyzers for clustering high-dimensional
  microarray data.
\newblock {\em Bioinformatics\/}~{\em 27\/}(9), 1269--1276.

\bibitem[\protect\citeauthoryear{Baek, McLachlan, and Flack}{Baek
  et~al.}{2010}]{baek10}
Baek, J., G.~J. McLachlan, and L.~K. Flack (2010).
\newblock {Mixtures of factor analyzers with common factor loadings:
  Applications to the clustering and visualization of high-dimensional data}.
\newblock {\em IEEE Transactions on Pattern Analysis and Machine
  Intelligence\/}~{\em 32\/}(7), 1298--1309.

\bibitem[\protect\citeauthoryear{Banfield and Raftery}{Banfield and
  Raftery}{1993}]{banfield93}
Banfield, J.~D. and A.~E. Raftery (1993).
\newblock Model-based {G}aussian and non-{G}aussian clustering.
\newblock {\em Biometrics\/}~{\em 49\/}(3), 803--821.

\bibitem[\protect\citeauthoryear{Barndorff-Nielsen and
  Bl{\ae}sild}{Barndorff-Nielsen and Bl{\ae}sild}{1981}]{blaesild81}
Barndorff-Nielsen, O. and P.~Bl{\ae}sild (1981).
\newblock Hyperbolic distributions and ramifications: Contributions to theory
  and application.
\newblock In C.~Taillie, G.~Patil, and B.~Baldessari (Eds.), {\em Statistical
  Distributions in Scientific Work}, Volume~79 of {\em NATO Advanced Study
  Institutes Series}, pp.\  19--44.

\bibitem[\protect\citeauthoryear{Barndorff-Nielsen and
  Halgreen}{Barndorff-Nielsen and Halgreen}{1977}]{barndorff77b}
Barndorff-Nielsen, O. and C.~Halgreen (1977).
\newblock {Infinite divisibility of the hyperbolic and generalized inverse
  {G}aussian distributions}.
\newblock {\em Probability Theory and Related Fields\/}~{\em 38\/}(4),
  309--311.

\bibitem[\protect\citeauthoryear{Bhattacharya and McNicholas}{Bhattacharya and
  McNicholas}{2014}]{bhattacharya14}
Bhattacharya, S. and P.~D. McNicholas (2014).
\newblock A {LASSO}-penalized {BIC} for mixture model selection.
\newblock {\em Advances in Data Analysis and Classification\/}~{\em 8\/}(1),
  45--61.

\bibitem[\protect\citeauthoryear{Bl{\ae}sild}{Bl{\ae}sild}{1978}]{blaesild78}
Bl{\ae}sild, P. (1978).
\newblock {\em {The Shape of the Generalized Inverse {G}aussian and Hyperbolic
  Distributions}}.
\newblock Department of Theoretical Statistics, Institute of Mathematics,
  University of Aarhus.

\bibitem[\protect\citeauthoryear{B\"ohning, Dietz, Schaub, Schlattmann, and
  Lindsay}{B\"ohning et~al.}{1994}]{bohning94}
B\"ohning, D., E.~Dietz, R.~Schaub, P.~Schlattmann, and B.~G. Lindsay (1994).
\newblock The distribution of the likelihood ratio for mixtures of densities
  from the one-parameter exponential family.
\newblock {\em Annals of the Institute of Statistical Mathematics\/}~{\em
  46\/}(2), 373--388.

\bibitem[\protect\citeauthoryear{Bouveyron and Brunet-Saumard}{Bouveyron and
  Brunet-Saumard}{2014}]{bouveyron14}
Bouveyron, C. and C.~Brunet-Saumard (2014).
\newblock {Model-based clustering of high-dimensional data: {A} review}.
\newblock {\em Computational Statistics and Data Analysis\/}~{\em 71}, 52--78.

\bibitem[\protect\citeauthoryear{Browne and McNicholas}{Browne and
  McNicholas}{2015}]{browne15}
Browne, R.~P. and P.~D. McNicholas (2015).
\newblock A mixture of generalized hyperbolic distributions.
\newblock {\em Canadian Journal of Statistics\/}~{\em 43\/}(2), 176--198.

\bibitem[\protect\citeauthoryear{Buuren and Groothuis-Oudshoorn}{Buuren and
  Groothuis-Oudshoorn}{2010}]{buuren10}
Buuren, S.~v. and K.~Groothuis-Oudshoorn (2010).
\newblock mice: {M}ultivariate imputation by chained equations in {R}.
\newblock {\em Journal of statistical software\/}, 1--68.

\bibitem[\protect\citeauthoryear{Celeux and Govaert}{Celeux and
  Govaert}{1995}]{celeux95}
Celeux, G. and G.~Govaert (1995).
\newblock Gaussian parsimonious clustering models.
\newblock {\em Pattern Recognition\/}~{\em 28\/}(5), 781--793.

\bibitem[\protect\citeauthoryear{Chi, Chi, and Baraniuk}{Chi
  et~al.}{2016}]{chi16}
Chi, J.~T., E.~C. Chi, and R.~G. Baraniuk (2016).
\newblock {k}-{POD}: A method for k-means clustering of missing data.
\newblock {\em The American Statistician\/}~{\em 70\/}(1), 91--99.

\bibitem[\protect\citeauthoryear{Dempster, Laird, and Rubin}{Dempster
  et~al.}{1977}]{dempster77}
Dempster, A.~P., N.~M. Laird, and D.~B. Rubin (1977).
\newblock {Maximum likelihood from incomplete data via the {EM} algorithm}.
\newblock {\em Journal of the Royal Statistical Society: Series~B\/}~{\em
  39\/}(1), 1--38.

\bibitem[\protect\citeauthoryear{Forina, Armanino, Castino, and Ubigli}{Forina
  et~al.}{1986}]{forina86}
Forina, M., C.~Armanino, M.~Castino, and M.~Ubigli (1986).
\newblock Multivariate data analysis as a discriminating method of the origin
  of wines.
\newblock {\em Vitis\/}~{\em 25\/}(3), 189--201.

\bibitem[\protect\citeauthoryear{Fraley and Raftery}{Fraley and
  Raftery}{2002}]{fraley02a}
Fraley, C. and A.~E. Raftery (2002).
\newblock Model-based clustering, discriminant analysis, and density
  estimation.
\newblock {\em Journal of the American Statistical Association\/}~{\em
  97\/}(458), 611--631.

\bibitem[\protect\citeauthoryear{Fraley, Raftery, Murphy, and Scrucca}{Fraley
  et~al.}{2012}]{fraley12}
Fraley, C., A.~E. Raftery, T.~B. Murphy, and L.~Scrucca (2012).
\newblock {\em mclust Version 4 for {R}: Normal Mixture Modeling for
  Model-Based Clustering, Classification, and Density Estimation}.

\bibitem[\protect\citeauthoryear{Ghahramani and Hinton}{Ghahramani and
  Hinton}{1997}]{ghahramani97}
Ghahramani, Z. and G.~E. Hinton (1997).
\newblock {The {EM} algorithm for mixtures of factor analyzers}.
\newblock Technical report, Technical Report CRG-TR-96-1, University of
  Toronto.

\bibitem[\protect\citeauthoryear{Good}{Good}{1953}]{good53}
Good, I.~J. (1953).
\newblock The population frequencies of species and the estimation of
  population parameters.
\newblock {\em Biometrika\/}~{\em 40\/}(3-4), 237--264.

\bibitem[\protect\citeauthoryear{H.}{H.}{2004}]{hurley04}
H., C.~B. (2004).
\newblock Clustering visualizations of multidimensional data.
\newblock {\em Journal of Computational and Graphical Statistics\/}~{\em
  13\/}(4), 788--806.

\bibitem[\protect\citeauthoryear{Halgreen}{Halgreen}{1979}]{halgreen79}
Halgreen, C. (1979).
\newblock {Self-decomposability of the generalized inverse {G}aussian and
  hyperbolic distributions}.
\newblock {\em Probability Theory and Related Fields\/}~{\em 47\/}(1), 13--17.

\bibitem[\protect\citeauthoryear{Honaker, King, Blackwell, et~al.}{Honaker
  et~al.}{2011}]{honaker11}
Honaker, J., G.~King, M.~Blackwell, et~al. (2011).
\newblock Amelia ii: A program for missing data.
\newblock {\em Journal of Statistical Software\/}~{\em 45\/}(7), 1--47.

\bibitem[\protect\citeauthoryear{Hubert and Arabie}{Hubert and
  Arabie}{1985}]{hubert85}
Hubert, L. and P.~Arabie (1985).
\newblock Comparing partitions.
\newblock {\em Journal of Classification\/}~{\em 2\/}(1), 193--218.

\bibitem[\protect\citeauthoryear{Hunter and Lange}{Hunter and
  Lange}{2000}]{hunter00}
Hunter, D.~L. and K.~Lange (2000).
\newblock Rejoinder to discussion of ``{O}ptimization transfer using surrogate
  objective functions''.
\newblock {\em Journal of Computational and Graphical Statistics\/}~{\em 9},
  52--59.

\bibitem[\protect\citeauthoryear{Hunter and Lange}{Hunter and
  Lange}{2004}]{hunter04}
Hunter, D.~L. and K.~Lange (2004).
\newblock A tutorial on {MM} algorithms.
\newblock {\em The American Statistician\/}~{\em 58\/}(1), 30--37.

\bibitem[\protect\citeauthoryear{J{\o}rgensen}{J{\o}rgensen}{1982}]{jorgensen82}
J{\o}rgensen, B. (1982).
\newblock {\em Statistical Properties of the Generalized Inverse {G}aussian
  Distribution}.
\newblock Lecture Notes in Statistics. New York: Springer.

\bibitem[\protect\citeauthoryear{Lawley and Maxwell}{Lawley and
  Maxwell}{1962}]{lawley62}
Lawley, D.~N. and A.~E. Maxwell (1962).
\newblock Factor analysis as a statistical method.
\newblock {\em Journal of the Royal Statistical Society. Series D (The
  Statistician)\/}~{\em 12\/}(3), 209--229.

\bibitem[\protect\citeauthoryear{Lichman}{Lichman}{2013}]{lichman13}
Lichman, M. (2013).
\newblock {UCI} machine learning repository.
\newblock University of California, Irvine, School of Information and Computer
  Sciences.

\bibitem[\protect\citeauthoryear{Lin, Ho, and Shen}{Lin et~al.}{2009}]{lin09b}
Lin, T.-I., H.~J. Ho, and P.~S. Shen (2009).
\newblock Computationally efficient learning of multivariate t mixture models
  with missing information.
\newblock {\em Computational Statistics\/}~{\em 24\/}(3), 375--392.

\bibitem[\protect\citeauthoryear{Lin, Lee, and Ho}{Lin et~al.}{2006}]{lin06}
Lin, T.-I., J.~C. Lee, and H.~J. Ho (2006).
\newblock On fast supervised learning for normal mixture models with missing
  information.
\newblock {\em Pattern Recognition\/}~{\em 39\/}(6), 1177--1187.

\bibitem[\protect\citeauthoryear{Lin, Lee, and Ni}{Lin et~al.}{2004}]{lin04}
Lin, T.-I., J.~C. Lee, and H.~F. Ni (2004).
\newblock Bayesian analysis of mixture modelling using the multivariate t
  distribution.
\newblock {\em Statistics and Computing\/}~{\em 14\/}(2), 119--130.

\bibitem[\protect\citeauthoryear{Lin, McLachlan, and Lee}{Lin
  et~al.}{2016}]{lin16}
Lin, T.-I., G.~J. McLachlan, and S.~Lee (2016).
\newblock Extending mixtures of factor models using the restricted multivariate
  skew-normal distribution.
\newblock {\em Journal of Multivariate Analysis\/}~{\em 143}, 398--413.

\bibitem[\protect\citeauthoryear{Lindsay}{Lindsay}{1995}]{lindsay95}
Lindsay, B.~G. (1995).
\newblock {Mixture Models: Theory, Geometry and Applications}.
\newblock In {\em {NSF-CBMS Regional Conference Series in Probability and
  Statistics}}, Volume~5. California: Institute of Mathematical Statistics:
  Hayward.

\bibitem[\protect\citeauthoryear{Little and Rubin}{Little and
  Rubin}{1987}]{little87}
Little, R.~J.~A. and D.~B. Rubin (1987).
\newblock {\em Statistical Analysis with Missing Data}.
\newblock New York: Wiley.

\bibitem[\protect\citeauthoryear{McLachlan and Peel}{McLachlan and
  Peel}{2000}]{mclachlan00}
McLachlan, G. and D.~Peel (2000).
\newblock {\em Finite Mixture Models}.
\newblock New York: Wiley.

\bibitem[\protect\citeauthoryear{McLachlan, Bean, and Jones}{McLachlan
  et~al.}{2007}]{mclachlan07}
McLachlan, G.~J., R.~W. Bean, and L.~B. Jones (2007).
\newblock Extension of the mixture of factor analyzers model to incorporate the
  multivariate t-distribution.
\newblock {\em Computational Statistics and Data Analysis\/}~{\em 51\/}(11),
  5327--5338.

\bibitem[\protect\citeauthoryear{McNeil, Frey, and Embrechts}{McNeil
  et~al.}{2005}]{mcneil05}
McNeil, A.~J., R.~Frey, and P.~Embrechts (2005).
\newblock {\em Quantitative Risk Management: Concepts, Techniques and Tools}.
\newblock Princeton, NJ: Princeton University Press.

\bibitem[\protect\citeauthoryear{McNicholas}{McNicholas}{2016a}]{mcnicholas16a}
McNicholas, P.~D. (2016a).
\newblock {\em Mixture Model-Based Classification}.
\newblock Boca Raton: Chapman \& Hall/CRC Press.

\bibitem[\protect\citeauthoryear{McNicholas}{McNicholas}{2016b}]{mcnicholas16b}
McNicholas, P.~D. (2016b).
\newblock Model-based clustering.
\newblock {\em Journal of Classification\/}~{\em 33\/}(3), 331--373.

\bibitem[\protect\citeauthoryear{McNicholas, ElSherbiny, McDaid, and
  Murphy}{McNicholas et~al.}{2015}]{mcnicholas15}
McNicholas, P.~D., A.~ElSherbiny, A.~F. McDaid, and T.~B. Murphy (2015).
\newblock {\em pgmm: Parsimonious {G}aussian Mixture Models}.
\newblock {R} package version 1.2.

\bibitem[\protect\citeauthoryear{McNicholas and Murphy}{McNicholas and
  Murphy}{2008}]{mcnicholas08}
McNicholas, P.~D. and T.~B. Murphy (2008).
\newblock {Parsimonious {G}aussian mixture models}.
\newblock {\em Statistics and Computing\/}~{\em 18\/}(3), 285--296.

\bibitem[\protect\citeauthoryear{McNicholas and Murphy}{McNicholas and
  Murphy}{2010}]{mcnicholas10}
McNicholas, P.~D. and T.~B. Murphy (2010).
\newblock {Model-based clustering of microarray expression data via latent
  {G}aussian mixture models}.
\newblock {\em Bioinformatics\/}~{\em 26\/}(21), 2705--2712.

\bibitem[\protect\citeauthoryear{McNicholas, Murphy, McDaid, and
  Frost}{McNicholas et~al.}{2010}]{mcnicholas10b}
McNicholas, P.~D., T.~B. Murphy, A.~F. McDaid, and D.~Frost (2010).
\newblock Serial and parallel implementations of model-based clustering via
  parsimonious {Gaussian} mixture models.
\newblock {\em Computational Statistics \& Data Analysis\/}~{\em 54\/}(3),
  711--723.

\bibitem[\protect\citeauthoryear{McNicholas, McNicholas, and Browne}{McNicholas
  et~al.}{2017}]{smcnicholas17}
McNicholas, S.~M., P.~D. McNicholas, and R.~P. Browne (2017).
\newblock A mixture of variance-{G}amma factor analyzers.
\newblock In S.~E. Ahmed (Ed.), {\em Big and Complex Data Analysis:
  Methodologies and Applications}, pp.\  369--385. Cham: Springer International
  Publishing.

\bibitem[\protect\citeauthoryear{Meng and Rubin}{Meng and Rubin}{1993}]{meng93}
Meng, X.-L. and D.~B. Rubin (1993).
\newblock {Maximum likelihood estimation via the ECM algorithm: A general
  framework}.
\newblock {\em Biometrika\/}~{\em 80\/}(2), 267--278.

\bibitem[\protect\citeauthoryear{Meng and Van~Dyk}{Meng and
  Van~Dyk}{1997}]{meng97}
Meng, X.-L. and D.~Van~Dyk (1997).
\newblock {The {EM} Algorithm---an Old Folk-song Sung to a Fast New Tune}.
\newblock {\em Journal of the Royal Statistical Society: Series B (Statistical
  Methodology)\/}~{\em 59\/}(3), 511--567.

\bibitem[\protect\citeauthoryear{Murray, Browne, and McNicholas}{Murray
  et~al.}{2014}]{murray14a}
Murray, P.~M., R.~P. Browne, and P.~D. McNicholas (2014).
\newblock Mixtures of skew-t factor analyzers.
\newblock {\em Computational Statistics and Data Analysis\/}~{\em 77},
  326--335.

\bibitem[\protect\citeauthoryear{{R Core Team}}{{R Core Team}}{2016}]{R16}
{R Core Team} (2016).
\newblock {\em R: {A} Language and Environment for Statistical Computing}.
\newblock Vienna, Austria: R Foundation for Statistical Computing.

\bibitem[\protect\citeauthoryear{Schwarz}{Schwarz}{1978}]{schwarz78}
Schwarz, G. (1978).
\newblock Estimating the dimension of a model.
\newblock {\em The Annals of Statistics\/}~{\em 6}, 461--464.

\bibitem[\protect\citeauthoryear{Su, Gelman, Hill, Yajima, et~al.}{Su
  et~al.}{2011}]{su11}
Su, Y., A.~Gelman, J.~Hill, M.~Yajima, et~al. (2011).
\newblock Multiple imputation with diagnostics (mi) in {R}: Opening windows
  into the black box.
\newblock {\em Journal of Statistical Software\/}~{\em 45\/}(2), 1--31.

\bibitem[\protect\citeauthoryear{Tortora, Browne, Franczak, and
  McNicholas}{Tortora et~al.}{2015}]{tortora15c}
Tortora, C., R.~P. Browne, B.~C. Franczak, and P.~D. McNicholas (2015).
\newblock {\em {MixGHD}: Model Based Clustering, Classification and
  Discriminant Analysis Using the Mixture of Generalized Hyperbolic
  Distributions}.
\newblock R package version 1.8.

\bibitem[\protect\citeauthoryear{Tortora, McNicholas, and Browne}{Tortora
  et~al.}{2016}]{tortora16}
Tortora, C., P.~D. McNicholas, and R.~P. Browne (2016).
\newblock A mixture of generalized hyperbolic factor analyzers.
\newblock {\em Advances in Data Analysis and Classification\/}~{\em 10\/}(4),
  423--440.

\bibitem[\protect\citeauthoryear{Utsugi and Kumagai}{Utsugi and
  Kumagai}{2001}]{utsugi01}
Utsugi, A. and T.~Kumagai (2001).
\newblock Bayesian analysis of mixtures of factor analyzers.
\newblock {\em Neural Computation\/}~{\em 13\/}(5), 993--1002.

\bibitem[\protect\citeauthoryear{V. and McNicholas}{V. and
  McNicholas}{2014}]{vrbik14}
Vrbik, I. and P.~D. McNicholas (2014).
\newblock Parsimonious skew mixture models for model-based clustering and
  classification.
\newblock {\em Computational Statistics and Data Analysis\/}~{\em 71},
  196--210.

\bibitem[\protect\citeauthoryear{Wagstaff and Laidler}{Wagstaff and
  Laidler}{2005}]{wagstaff05}
Wagstaff, K.~L. and V.~G. Laidler (2005).
\newblock Making the most of missing values: Object clustering with partial
  data in astronomy.
\newblock In {\em Astronomical Data Analysis Software and Systems XIV}, Volume
  347, pp.\  172.

\bibitem[\protect\citeauthoryear{Wang}{Wang}{2013}]{wang13}
Wang, W.-L. (2013).
\newblock Mixtures of common factor analyzers for high-dimensional data with
  missing information.
\newblock {\em Journal of Multivariate Analysis\/}~{\em 117}, 120 --133.

\bibitem[\protect\citeauthoryear{Wang}{Wang}{2015}]{wang15}
Wang, W.-L. (2015).
\newblock Mixtures of common t-factor analyzers for modeling high-dimensional
  data with missing values.
\newblock {\em Computational Statistics and Data Analysis\/}~{\em 83},
  223--235.

\bibitem[\protect\citeauthoryear{Wei, Tang, and McNicholas}{Wei
  et~al.}{2019}]{wei17}
Wei, Y., Y.~Tang, and P.~D. McNicholas (2019).
\newblock Mixtures of generalized hyperbolic distributions and mixtures of
  skew-t distributions for model-based clustering with incomplete data.
\newblock {\em Computational Statistics and Data Analysis\/}~{\em 130}, 
  18--41.

\bibitem[\protect\citeauthoryear{Zhang and Fan}{Zhang and Fan}{2008}]{zhang08}
Zhang, K. and W.~Fan (2008).
\newblock Forecasting skewed biased stochastic ozone days: analyses, solutions
  and beyond.
\newblock {\em Knowledge and Information Systems\/}~{\em 14\/}(3), 299--326.

\bibitem[\protect\citeauthoryear{Zhang, Fan, Yuan, Davidson, and Li}{Zhang
  et~al.}{2006}]{zhang06}
Zhang, K., W.~Fan, X.~Yuan, I.~Davidson, and X.~Li (2006).
\newblock Forecasting skewed biased stochastic ozone days: Analyses and
  solutions.
\newblock In {\em Proceedings of the Sixth International Conference on Data
  Mining}, pp.\  753--764. IEEE.

\end{thebibliography}
\end{document}